\documentclass[aps,showpacs,floatfix]{revtex4}
\usepackage{amssymb}
\usepackage{amsmath}
\usepackage{epsfig}
\usepackage{subfigure}

\begin{document}

\title{The Power of General Relativity }
\author{Timothy Clifton}
\email{T.Clifton@damtp.cam.ac.uk}
\affiliation{DAMTP, Centre for Mathematical Sciences, University of Cambridge,
Wilberforce Road, Cambridge, CB3 0WA, UK}
\author{John D. Barrow}
\email{J.D.Barrow@damtp.cam.ac.uk}
\affiliation{DAMTP, Centre for Mathematical Sciences, University of Cambridge,
Wilberforce Road, Cambridge, CB3 0WA, UK}
\date{\today }

\begin{abstract}
We study the cosmological and weak-field properties of theories of gravity
derived by extending general relativity by means of a Lagrangian
proportional to $R^{1+\delta }$. This scale-free extension reduces to
general relativity when $\delta \rightarrow 0$. In order to constrain
generalisations of general relativity of this power class we analyse the
behaviour of the perfect-fluid Friedmann universes and isolate the
physically relevant models of zero curvature. A stable matter-dominated
period of evolution requires $\delta >0$ or $\delta <-1/4$. The stable
attractors of the evolution are found. By considering the synthesis of light
elements (helium-4, deuterium and lithium-7) we obtain the bound $%
-0.017<\delta <0.0012.$ We evaluate the effect on the power spectrum of
clustering via the shift in the epoch of matter-radiation equality. The
horizon size at matter--radiation equality will be shifted by $\sim 1\%$ for
a value of $\delta \sim 0.0005.$ We study the stable extensions of the
Schwarzschild solution in these theories and calculate the timelike and null
geodesics. No significant bounds arise from null geodesic effects but the
perihelion precession observations lead to the strong bound $\delta =2.7\pm
4.5\times 10^{-19}$ assuming that Mercury follows a timelike geodesic. The
combination of these observational constraints leads to the overall bound $%
0\leq \delta <7.2\times 10^{-19}$ on theories of this type.
\end{abstract}

\maketitle

\section{Introduction}

There is a long history of considering generalisations of Einstein's theory
of general relativity which reduce to general relativity in the weak gravity
limit when the spacetime curvature, $R$, becomes small. Typically, these
studies consider a gravitational Lagrangian which augments the linear
Einstein-Hilbert Lagrangian by the addition of terms of quadratic or higher
order in $R,$ first considered by Eddington \cite{edd}; these additions may
also include terms in $\ln R$, \cite{gur}. More general extensions of
general relativity in this spirit have considered the structure of theories
derived from gravitational Lagrangians that are general analytic functions
of $R$, \cite{buch, kerner, BO, magg}. These choices produce theories which
can look like general relativity plus small polynomial corrections in the
appropriate limiting situations as $R$ becomes small. There has also been
interest in theories with corrections to general relativity that are $%
O(R^{-1})$ because of their scope to introduce cosmological deviations from
general relativity at late times which might mimic the effects of dark
energy on the Hubble flow \cite{tur, new}. We also know that theories derived
from a Lagrangian that is an analytic function of $R$ have an important
conformal relationship to general relativity with scalar field sources so
long as the trace of the energy-momentum tensor vanishes in the higher-order
gravity theory \cite{CB, maeda}. All these theories introduce corrections to
general relativity which come with a characteristic length scale that is
determined by the new coupling constant that couples the higher-order terms
to the Einstein-Hilbert part of the Lagrangian. In general, these theories
are mathematically complicated with 4$^{th}$-order field equations that can
exhibit singular perturbation behaviour unless care is taken to ensure that
the stationary action does not become maximal rather than minimal \cite{ruz,
BO}. and there are few interesting exact solutions other than those of
general relativity, which are particular solutions in vacuum and for
trace-free energy momentum tensor so long as the cosmological constant
vanishes \cite{BO}.

In this paper we are going to consider a different type of generalisation of
Einstein's general relativity, in which no new scale is introduced. The
Lagrangian is proportional to $R^{n}$, and so general relativity is
recovered in the $n\rightarrow 1$ limit, from above or below. Particular
cases have been studied by Buchdahl \cite{Buc70} and Roxburgh \cite{rox}.
This gravitation theory has many appealing properties and, unlike
other higher-order gravity theories, admits simple
exact solutions for Friedmann cosmological models and exact static
spherically symmetric solutions which generalise the Schwarzschild
metric.  As well as allowing comparison with observation these
solutions also provide an interesting testing
ground for new developments in gravitation theory such as particle production,
holography and gravitational thermodynamics.  Furthermore, 
this theory is of additional interest because it permits a very general
investigation of the nature of its behaviour in the vicinity of a
cosmological singularity which brings the behaviour of general
relativity into sharper focus.  In another paper \cite{Bar05}, we show
that the counterpart of the Kasner anisotropic vacuum cosmology can be
found exactly and strong conclusions drawn about the presence or absence
of the chaotic behaviour found in the Mixmaster universe.

The structure of this paper is as follows; in the next section we present
the gravitational action and field equations for the theory of gravity we
will be considering. A conformal relationship with general relativity
containing a scalar field in a Liouville (exponential) potential is then
outlined and the Newtonian limit of the field equations is investigated. The
rest of the paper is then split into two sections; the first investigates
the cosmology of the theory and the second investigates the static and
spherically symmetric weak field - in both cases our goal is to calculate
predictions for physical processes, the results of which can be compared
with observation. We use observational data from cosmology and the standard
solar system tests of general relativity to bound the allowed values of $n$,
the single defining parameter of the theory.

In the cosmology section we consider Friedmann--Robertson--Walker universes.
We present the equivalent of the Friedmann equations, in this theory, and
find some power--law exact solutions. A dynamical systems approach is then
used to show the extent to which these solutions can be considered as
attractors of spatially flat universes at late times. After showing the
attractive properties of these solutions (with certain exceptions) we
proceed to predict the results of primordial nucleosynthesis and the form of
the power spectrum of perturbations in this theory. These predictions are
then compared to observation and used to constrain deviations from general
relativity.

The static and spherically symmetric weak-field analysis follows. We present
the field equations and find the physically relevant exact solution to them.
A dynamical systems approach is then used to find the asymptotic attractor
of the general solution at large distances. This asymptotic form is then
perturbed and the linearised field equations are found and solved. The exact
solution is shown to be the relevant solution in this limit, when
oscillatory modes in the perturbed metric functions are set to zero. We find
the null and time-like geodesics for this spacetime to Newtonian and
post-Newtonian order. Predictions are then made for the outcomes of the
classical tests of general relativity in this theory; namely the bending of
light, the time-delay of radio signals and the perihelion precession of
Mercury. These predictions are then compared to observation and again used
to constrain deviations from general relativity.

\section{Field equations}

We consider here a gravitational theory derived from the Lagrangian density 
\begin{equation}
\mathcal{L}_{G}=\frac{1}{\chi }\sqrt{-g}R^{1+\delta },  \label{action}
\end{equation}%
where $\delta $ is a real number and $\chi $ is a constant. The limit $%
\delta \rightarrow 0$ gives us the familiar Einstein--Hilbert Lagrangian of
general relativity and we are interested in the observational consequences
of $\left\vert \delta \right\vert $ $>0$.

We denote the matter action as $S_{m}$ and ignore the boundary term.
Extremizing 
\begin{equation*}
S=\int \mathcal{L}_{G}d^{4}x+S_{m},
\end{equation*}%
with respect to the metric $g_{ab}$ then gives \cite{Buc70} 
\begin{multline}
\delta (1-\delta ^{2})R^{\delta }\frac{R_{,a}R_{,b}}{R^{2}}-\delta (1+\delta
)R^{\delta }\frac{R_{;ab}}{R}+(1+\delta )R^{\delta }R_{ab}-\frac{1}{2}%
g_{ab}RR^{\delta }  \label{field} \\
-g_{ab}\delta (1-\delta ^{2})R^{\delta }\frac{R_{,c}R_{,}^{\ c}}{R^{2}}%
+\delta (1+\delta )g_{ab}R^{\delta }\frac{\Box R}{R}=\frac{\chi }{2}T_{ab},
\end{multline}%
where $T_{ab}$ is the energy--momentum tensor of the matter, and is defined
in terms of $S_{m}$ and $g_{ab}$ in the usual way. We take the quantity $%
R^{\delta }$ to be the positive real root of $R$ throughout this paper.

\subsection{Conformal equivalence to general relativity}

Rescaling the metric by the conformal factor $\Omega (r)=\Omega
_{0}R^{\delta }$ the vacuum field equations (\ref{field}) become 
\begin{equation*}
\bar{G}_{ab}=\frac{3\delta ^{2}}{2}\frac{R_{,a}R_{,b}}{R^{2}}-\frac{3\delta
^{2}}{4}\bar{g}_{ab}\bar{g}^{cd}\frac{R_{,c}R_{,d}}{R^{2}}-\frac{\delta }{%
2(1+\delta )}\frac{\bar{g}_{ab}}{\Omega _{0}}\frac{R}{R^{\delta }},
\end{equation*}%
where $\bar{g}_{ab}=\Omega g_{ab}$ and other quantities with overbars are
constructed from the rescaled metric $\bar{g}_{ab}$.

Making the definition of a scalar field

\begin{equation*}
\phi \equiv \sqrt{\frac{3}{16\pi G}}\ln R^{\delta },
\end{equation*}

these equations can be rewritten as 
\begin{equation}
\bar{G}_{ab}=8\pi G\left( \phi _{,a}\phi _{,b}-\frac{1}{2}\bar{g}_{ab}(\bar{g%
}^{cd}\phi _{,c}\phi _{,d}+2 V(\phi ))\right)  \label{conformalfield}
\end{equation}%
and 
\begin{equation*}
\square \phi =\frac{dV}{d\phi },
\end{equation*}%
where $V(\phi )$ is given by 
\begin{equation}
V(\phi )=\frac{\delta \ \text{sign}(R)}{16\pi G(1+\delta )\Omega _{0}}\exp
\left\{ {\sqrt{\frac{16\pi G}{3}}\frac{(1-\delta )}{\delta }\phi }\right\} .
\label{pot1}
\end{equation}%
The magnitude of the quantity $\Omega _{0}$ is not physically important and
simply corresponds to the rescaling of the metric by a constant quantity,
which can be absorbed by an appropriate rescaling of units. It is, however,
important to ensure that $\Omega _{0}>0$ in order to maintain the +2
signature of the metric. This result is a particular example of the general
conformal equivalence to general relativity plus a scalar field for
Lagrangians of the form $f(R)$, where $f$ is an analytic function found in
refs. \cite{CB, maeda}.

\subsection{The Newtonian Limit}

By comparing the geodesic equation to Newton's gravitational force law it
can be seen that, as usual, 
\begin{equation}
\Gamma _{00}^{\mu }=\Phi _{,\mu }  \label{ChrN}
\end{equation}%
where $\Phi $ is the Newtonian gravitational potential. All the other
Christoffel symbols have $\Gamma _{\;bc}^{a}=0$, to the required order of
accuracy.

We now seek an approximation to the field equations (\ref{field}) that is of
the form of Poisson's equation; this will allow us to fix the constant $\chi 
$. Constructing the components of the Riemann tensor from (\ref{ChrN}) we
obtain the standard results 
\begin{equation}
R_{\;0\nu 0}^{\mu }=\frac{\partial ^{2}\Phi }{\partial x^{\mu }\partial
x^{\nu }}\qquad \text{and}\qquad R_{00}=\nabla ^{2}\Phi .  \label{R00}
\end{equation}%
The $00$ component of the field equations (\ref{field}) can now be written 
\begin{equation}
(1+\delta )R_{00}-\frac{1}{2}g_{00}R=\frac{\chi }{2}\frac{T_{00}}{R^{\delta }%
}  \label{approx}
\end{equation}%
where terms containing derivatives of $R$ have been discarded as they will
contain third and fourth derivatives of $\Phi $, which will have no
counterparts in Poisson's equation. Subtracting the trace of equation (\ref%
{approx}) gives 
\begin{equation}
(1+\delta )R_{00}=\frac{\chi }{2R^{\delta }}\left( T_{00}-\frac{1}{%
2(1-\delta )}g_{00}T\right)  \label{approx2}
\end{equation}%
where $T$ is the trace of the stress--energy tensor. Assuming a
perfect--fluid form for $T$ we should have, to first--order, 
\begin{equation}
T_{00}\simeq \rho \qquad \text{and}\qquad T\simeq 3p-\rho \simeq -\rho .
\label{T00}
\end{equation}%
Substituting (\ref{T00}) and (\ref{R00}) into (\ref{approx2}) gives 
\begin{equation*}
\nabla ^{2}\Phi \simeq \frac{\chi (1-2\delta )}{4(1-\delta ^{2})}\frac{\rho 
}{R^{\delta }}.
\end{equation*}%
Comparison of this expression with Poisson's equation allows one to read off 
\begin{equation}
\chi =16\pi G\frac{(1-\delta ^{2})}{(1-2\delta )}R_{0}^{\delta }  \label{chi}
\end{equation}%
where $R_{0}$ is the value of the Ricci tensor at the time $G$ is measured.
It can be seen that the Newtonian limit of the field equations (\ref{field})
does not reduce to the usual relation $\nabla ^{2}\Phi \propto \rho $, but
instead contains an extra factor of $R^{\delta }$. This can be interpreted
as being the space--time dependence of Newton's constant, in this theory.
Such a dependence should be expected as the Lagrangian (\ref{action}) can be
shown to be equivalent to a scalar--tensor theory, after an appropriate
Legendre transformation \footnote{This equivalence to
scalar-tensor theories should not be taken to imply that bounds on the
Brans-Dicke parameter $\omega$ are immediately applicable to this
theory.  It can be shown that a potential for the scalar-field can
have a non-trivial effect on the resulting phenomenology of the theory
\cite{Olmo}.  Furthermore, the form of the perturbation to general
relativity that we are considering does not allow an expansion of the
corresponding scalar field of the form $\phi_0+\phi_1$ where $\phi_0$ is constant
and $\vert \phi_1 \vert << \vert \phi_0 \vert$, so that any
constraints obtained in a weak-field expansion of this sort cannot be
applied to this situation.} (see e.g, \cite{Mag94}).  This type of Newtonian
gravity theory admits a range of simple exact solutions in the case where
the effective value of $G$ is a power-law in time even though the theory
is non-conservative and there is no longer an energy integral
\cite{newt}.

\section{Cosmology}

In this paper we will be concerned with the idealised homogeneous and
isotropic space--times described by the Friedmann--Robertson--Walker metric
with curvature parameter $\kappa $: 
\begin{equation}
ds^{2}=-dt^{2}+a^{2}(t)\left( \frac{dr^{2}}{(1-\kappa r^{2})}+r^{2}d\theta
^{2}+r^{2}\sin ^{2}\theta d\phi ^{2}\right) .  \label{FRW}
\end{equation}%
Substituting this metric ansatz into the field equations (\ref{field}), and
assuming the universe to be filled with a perfect fluid of pressure $p$ and
density $\rho $, gives the generalised version of the Friedmann equations 
\begin{align}
(1-\delta )R^{1+\delta }+3\delta (1+\delta )R^{\delta }\left( \frac{\ddot{R}%
}{R}+3\frac{\dot{a}}{a}\frac{\dot{R}}{R}\right) -3\delta (1-\delta
^{2})R^{\delta }\frac{\dot{R}^{2}}{R^{2}}& =\frac{\chi }{2}(\rho -3p)
\label{Friedman1} \\
-3\frac{\ddot{a}}{a}(1+\delta )R^{\delta }+\frac{R^{1+\delta }}{2}+3\delta
(1+\delta )\frac{\dot{a}}{a}\frac{\dot{R}}{R}R^{\delta }& =\frac{\chi }{2}%
\rho
\label{Friedman2}
\end{align}%
where, as usual, 
\begin{equation}
R=6\frac{\ddot{a}}{a}+6\frac{\dot{a}^{2}}{a^{2}}+6\frac{\kappa }{a^{2}}.
\label{R2}
\end{equation}%
It can be seen that in the limit $\delta \rightarrow 0$ these equations
reduce to the standard Friedmann equations of general relativity. A study of
the vacuum solutions to these equations for all $\kappa $ has been made by
Schmidt, see the review \cite{schmidt} and a qualitative study of the
perfect-fluid evolution for all $\kappa $ has been made by Carloni et al 
\cite{Car04}. Various conclusions are also immediate from the general
analysis of $f(R)$ Lagrangians made in ref \cite{BO} by specialising them to
the case $f=R^{1+\delta }$. In what follows we shall be interested in
extracting the physically relevant aspects of the general evolution so that
observational bounds can be placed on the allowed values of $\delta $.

Assuming a perfect-fluid equation of state of the form $p=\omega \rho $
gives the usual conservation equation $\rho \propto a^{-3(\omega +1)}$.
Substituting this into equations (\ref{Friedman1}) and (\ref{Friedman2}),
with $\kappa =0$, gives the power--law exact Friedmann solution for $\omega
\neq -1$ 
\begin{equation}
a(t)=t^{\frac{2(1+\delta )}{3(1+\omega )}}  \label{power}
\end{equation}%
where 
\begin{equation}
(1-2\delta )(2-3\delta (1+\omega )-2\delta ^{2}(4+3\omega ))=12\pi
G(1-\delta )(1+\omega )^{2}\rho _{c}  \label{rhoc}
\end{equation}%
and $\rho _{c}$ is the critical density of the universe.

Alternatively, if $\omega =-1$, there exists the de Sitter solution 
\begin{equation*}
a(t)=e^{nt}
\end{equation*}%
where

\begin{equation*}
3(1-2\delta )n^{2}=8\pi G(1-\delta )\rho _{c}.
\end{equation*}

The critical density (\ref{rhoc}) is shown graphically, in figure \ref%
{density}, in terms of the density parameter $\Omega _{0}=\frac{8\pi G\rho
_{c}}{3H_{0}^{2}}$ as a function of $\delta $ for pressureless dust ($\omega
=0$) and black-body radiation ($\omega =1/3$). It can be seen from the graph
that the density of matter required for a flat universe is dramatically
reduced for positive $\delta $, or large negative $\delta $. In order for
the critical density to correspond to a positive matter density we require $%
\delta $ to lie in the range 
\begin{equation}
-\frac{\sqrt{73+66\omega +9\omega ^{2}}+3(1+\omega )}{4(4+3\omega )}<\delta <%
\frac{\sqrt{73+66\omega +9\omega ^{2}}-3(1+\omega )}{4(4+3\omega )}.
\label{range}
\end{equation}

\begin{figure}[tbp]
\epsfig{file=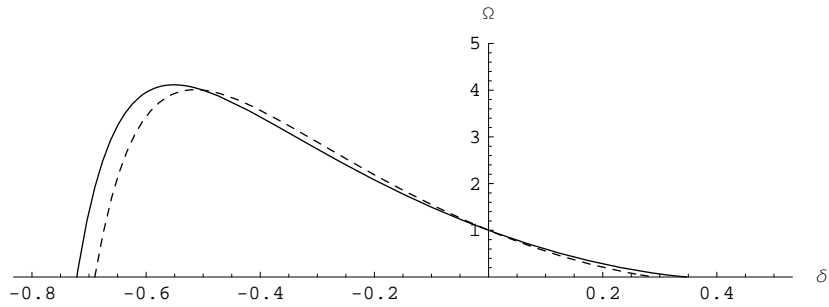,height=4cm}
\caption{\textit{Critical density, $\Omega_0$, as a function of $\protect%
\delta$. Solid line corresponds to pressure-less dust and dashed line to
black--body radiation.}}
\label{density}
\end{figure}

\subsection{The dynamical systems approach}

The system of equations (\ref{Friedman1}) and (\ref{Friedman2}) have been
studied previously using a dynamical systems approach by Carloni, Dunsby,
Capozziello and Troisi for general $\kappa $ \cite{Car04}. We elaborate on
their work by studying in detail the spatially flat, $\kappa =0$, subspace
of solutions. This allows us to draw conclusions about the asymptotic
solutions of (\ref{Friedman1}) and (\ref{Friedman2}) when $\kappa =0$ and so
investigate the stability of the power--law exact solution (\ref{power}) and
the extent to which it can be considered an attractor solution. By
restricting to $\kappa =0$ we avoid `instabilities' associated with the
curvature which are already present in general relativistic cosmologies.

In performing this analysis we choose to work in the conformal time
coordinate 
\begin{equation}  \label{tau}
d \tau \equiv \sqrt{\frac{8 \pi \rho}{3 R^{\delta}}} dt.
\end{equation}

Making the definitions 
\begin{equation*}
x \equiv \frac{R^{\prime}}{R} \qquad \text{and} \qquad y \equiv \frac{%
a^{\prime}}{a},
\end{equation*}
where a prime indicates differentiation with respect to $\tau$, the field
equations (\ref{Friedman1}) and (\ref{Friedman2}) can be written as the
autonomous set of first order equations 
\begin{align}  \label{phase1}
x^{\prime}&= \frac{2- \delta (1+ 3 \omega)}{\delta^2 (1+\delta)}- \frac{%
\delta x^2}{2}-\frac{(4-\delta (1+3 \omega)) x y}{2 \delta}- \frac{2
(1-\delta) y^2}{\delta^2} \\
y^{\prime}&= -\frac{1}{\delta}+\frac{1}{2} (2+3 \delta) x y+\frac{(2+\delta
(1+3 \omega)) y^2}{2 \delta}.
\label{phase2}
\end{align}

These coordinate definitions are closely related to those chosen by Holden
and Wands \cite{Hol98} for their phase-plane analysis of Brans-Dicke
cosmologies and allow us to proceed in a similar fashion.

\subsubsection{Locating the critical points}

The critical points at finite distances in the system of equations (\ref%
{phase1}) and (\ref{phase2}) are located at 
\begin{equation}  \label{critical1}
x_{1,2} = \pm \frac{1-3 \omega}{\delta \sqrt{(1+ \delta) (2-3 \omega)}}
\qquad \text{and} \qquad y_{1,2} = \pm \frac{1}{\sqrt{(1+\delta) (2-3 \omega)%
}}
\end{equation}
and at 
\begin{equation}  \label{critical2}
x_{3,4} = \mp \frac{3 \sqrt{2 (1+\omega)}}{\sqrt{(1+\delta) (2-3 \delta
(1+\omega)-\delta^2 (8+6 \omega))}} \qquad \text{and} \qquad y_{3,4} = \pm 
\frac{\sqrt{2 (1+\delta)}}{\sqrt{2-3 \delta (1+\omega)-\delta^2 (8+6 \omega)}%
}.
\end{equation}

The exact form of $a(t)$ at these critical points, and the stability of
these solutions, can be easily deduced. At the critical point $(x_{i},y_{i})$
the forms of $a(\tau )$ and $R(\tau )$ are given by 
\begin{equation}
a(\tau )=a_{0}e^{y_{i}\tau }\qquad \text{and}\qquad R(\tau
)=R_{0}e^{x_{i}\tau },  \label{atau}
\end{equation}%
where $a_{0}$ and $R_{0}$ are constants of integration. In terms of $\tau $
the perfect-fluid conservation equation can be integrated to give 
\begin{equation*}
\rho =\rho _{0}e^{-3(1+\omega )y_{i}\tau },
\end{equation*}%
where $\rho _{0}$ is another positive constant. Substituting into the
definition of $\tau $ now gives 
\begin{equation*}
d\tau \propto e^{-\frac{3}{2}(1+\omega )y_{i}\tau -\frac{\delta }{2}%
x_{i}\tau }dt
\end{equation*}%
or, integrating, 
\begin{equation}
t-t_{0}\propto \frac{1}{\frac{3}{2}(1+\omega )y_{i}+\frac{\delta }{2}x_{i}}%
e^{\frac{3}{2}(1+\omega )y_{i}\tau +\frac{\delta }{2}x_{i}\tau }.  \label{t}
\end{equation}

It can now be seen that if $3 (1+\omega) y_i+\delta x_i >0$ then $t
\rightarrow \infty$ as $\tau \rightarrow \infty$ and $t \rightarrow t_0$ as $%
\tau \rightarrow -\infty$. Conversely, if $3 (1+\omega) y_i+\delta x_i <0$
then $t \rightarrow t_0$ as $\tau \rightarrow \infty$ and $t \rightarrow
-\infty$ as $\tau \rightarrow -\infty$.

In terms of $t$ time the solutions corresponding to the critical points at
finite distances can now be written as 
\begin{equation*}
a(t)\propto (t-t_{0})^{\frac{2y_{i}}{3(1+\omega )y_{i}+\delta x_{i}}}\qquad 
\text{and}\qquad R(t)\propto (t-t_{0})^{\frac{2x_{i}}{3(1+\omega
)y_{i}+\delta x_{i}}}.
\end{equation*}%
The critical points 1 and 2 can now been seen to correspond to $a\propto t^{%
\frac{1}{2}}$ and the points 3 and 4 correspond to (\ref{power}).

In order to analyse the behaviour of the solutions as they approach infinity
it is convenient to transform to the polar coordinates 
\begin{align*}
x &= \bar{r} \cos \phi \\
y &= \bar{r} \sin \phi.
\end{align*}
The infinite phase plane can then be compacted into a finite size by
introducing the coordinate 
\begin{equation*}
r = \frac{\bar{r}}{1+\bar{r}}.
\end{equation*}
The equations (\ref{phase1}) and (\ref{phase2}) then become 
\begin{multline}  \label{phase3}
r^{\prime}=\frac{-1}{4 \delta^2 (1+\delta)} \Biggl( 4 (1-2 r) (\delta
(1+\delta) \sin \phi-(2-\delta (1+3 \omega)) \cos \phi) \\
- r^2 ((6-4 \delta+3 \delta^2+\delta^3-12 \delta \omega) \cos \phi
+(1+\delta) (2-2 \delta-\delta^2-2 \delta^3) \cos 3 \phi \\
- 2 \delta (3-\delta (1+3 \omega)+3 \cos 2 \phi) \sin \phi) \Biggr)
\end{multline}
and 
\begin{multline}  \label{phase4}
\phi^{\prime}=\frac{-1}{2 \delta^2 (1+\delta) (1-r) r} \Biggl( (2 \delta
(1+\delta) \cos \phi+2(2-\delta (1+3 \omega)) \sin \phi) (1-2 r) \\
-\Bigl( \delta (1+\delta) \cos \phi (1-3 \cos 2 \phi)-4 \sin \phi+4
(1-\delta)^2 \sin^3 \phi \\
+2 \delta (1+3 \omega+\delta (1+\delta) (1+2 \delta) \cos^2 \phi) \sin \phi
\Bigr) r^2 \Biggr).
\end{multline}

In the limit $r\rightarrow 1$ ($\bar{r} \rightarrow \infty$) it can be seen
that critical points at infinity satisfy 
\begin{equation*}
\sin \phi_i (\delta \cos \phi_i+\sin \phi_i) (\delta (1+2 \delta) \cos
\phi_i+2 (1-\delta) \sin \phi_i)=0
\end{equation*}
and so are located at 
\begin{align}  \label{5,6}
\phi_{5,(6)} &=0\qquad(+\pi) \\
 \label{7,8}
\phi_{7,(8)} &=\tan^{-1} (-\delta)\qquad (+\pi) \\
\phi_{9,(10)} &=\tan^{-1} \left( -\frac{\delta (1+2 \delta)}{2 (1-\delta)}%
\right) \qquad(+\pi).
 \label{9,10}
\end{align}
The form of $a(t)$ can now be calculated for each of these critical points
by proceeding as Holden and Wands \cite{Hol98}. Firstly, as $r \rightarrow 1$
equation (\ref{phase3}) approaches 
\begin{align*}
r^{\prime}&\rightarrow \frac{1}{4 \delta^2} \Biggl( \delta (1+2 \delta (1+3
\omega)) \sin \phi_i-3 \delta \sin 3 \phi_i \\
&\qquad -(2-\delta (2+\delta))\cos \phi_i+(2-\delta (2+\delta+2 \delta^2))
\cos 3 \phi_i\Biggr) \\
&\equiv f(\phi_i)
\end{align*}
which allows the integral 
\begin{equation*}
r-1= f(\phi_i) (\tau-\tau_0)
\end{equation*}
where the constant of integration, $\tau_0$, has been set so that $r
\rightarrow 1$ as $\tau \rightarrow \tau_0$. Now the definition of $x$
allows us to write 
\begin{equation*}
\frac{R^{\prime}}{R}=\frac{r}{(1-r)} \cos \phi_i=-\frac{f(\phi_i)
(\tau-\tau_0)+1}{f(\phi_i) (\tau-\tau_0)} \cos \phi_i \rightarrow -\frac{%
\cos \phi_i}{f(\phi_i) (\tau-\tau_0)}
\end{equation*}
as $\tau \rightarrow \tau_0$. Integrating this it can be seen that 
\begin{equation*}
R \propto \vert \tau-\tau_0 \vert^{-\frac{\cos \phi_i}{f(\phi_i)}} \qquad 
\text{as} \qquad r \rightarrow 1.
\end{equation*}
Similarly, 
\begin{equation*}
a \propto \vert \tau-\tau_0 \vert^{-\frac{\sin \phi_i}{f(\phi_i)}} \qquad 
\text{as} \qquad r \rightarrow 1.
\end{equation*}

The definition of $\tau$ (\ref{tau}) now gives 
\begin{equation*}
d \tau \propto \vert \tau-\tau_0 \vert^{\frac{3}{2} (1+\omega) \frac{%
\sin\phi_i}{f(\phi_i)}+\frac{\delta}{2} \frac{\cos\phi_i}{f(\phi_i)}} dt
\end{equation*}
which integrates to 
\begin{equation}  \label{tinfty}
t-t_0 \propto - \frac{f(\phi_1)}{F(\phi_i)} \vert \tau-\tau_0 \vert^{-\frac{%
F(\phi_i)}{f(\phi_i)}}
\end{equation}
where 
\begin{equation*}
F(\phi_i)=\frac{3 (1+\omega) \sin \phi_i +\delta \cos \phi_i-2 f(\phi_i)}{2}.
\end{equation*}
The location of critical points at infinity can now be written in terms of $%
t $ as the power--law solutions 
\begin{equation}  \label{inftysol}
R(t) \propto (t-t_0)^{\frac{\cos\phi_i}{F(\phi_i)}} \qquad \text{and} \qquad
a(t) \propto (t-t_0)^{\frac{\sin\phi_i}{F(\phi_i)}}.
\end{equation}

Direct substitution of the critical points (\ref{5,6}), (\ref{7,8}) and (\ref%
{9,10}) into (\ref{inftysol}) gives 
\begin{align*}
a_{5,6}(t)& \rightarrow \ constant \\
a_{7,8}(t)& \rightarrow \sqrt{t-t_{0}} \\
a_{9,10}(t)& \rightarrow (t-t_{0})^{\frac{\delta (1+2\delta )}{(1-\delta )}}
\end{align*}%
as $r\rightarrow 1$. Moreover, it can be seen from (\ref{tinfty}) that as $%
r\rightarrow 1$ and $\tau \rightarrow \tau _{0}$ so $t\rightarrow t_{0}$ as
long as $F(\phi _{i})/f(\phi _{i})<0$, as is the case for the stationary
points considered here (as long as the value of $\delta $ lies within the
range given by (\ref{range})).

The exact forms of $a(t)$ at all the critical points are summarised in the
table below.

\begin{center}
\begin{tabular}{c|c}
\textbf{Critical point} & \textbf{a(t)} \\ \hline
1, 2, 7 and 8 & $t^{\frac{1}{2}}$ \\ 
3 and 4 & $t^{\frac{2 (1+\delta)}{3 (1+\omega)}}$ \\ 
5 and 6 & constant \\ 
9 and 10 & $t^{\frac{\delta (1+2 \delta)}{(1-\delta)}}$%
\end{tabular}
\end{center}

\subsubsection{Stability of the critical points}

The stability of the critical points at finite distances can be established
by perturbing $x$ and $y$ as 
\begin{equation}
x(r)=x_{i}+u(r)\qquad \text{and}\qquad y(r)=y_{i}+v(r)  \label{lin}
\end{equation}%
and checking the sign of the eigenvalues, $\lambda _{i}$, of the linearised
equations 
\begin{equation*}
u^{\prime }=\lambda _{i}u\qquad \text{and}\qquad v^{\prime }=\lambda _{i}v.
\end{equation*}

Substituting (\ref{lin}) into equations (\ref{phase1}) and (\ref{phase2})
and linearising in $u$ and $v$ gives 
\begin{align*}
u^{\prime }& =-\left( \delta x_{i}+\frac{(4-\delta (1+3\omega ))}{2\delta }%
y_{i}\right) u-\left( \frac{(4-\delta (1+3\omega ))}{2\delta }x_{i}+4\frac{%
(1+\delta )}{\delta ^{2}}y_{i}\right) v \\
v^{\prime }& =\frac{(2+3\delta )}{2}y_{0}u+\left( \frac{(2+3\delta )}{2}%
x_{i}+\frac{(2+\delta (1+3\omega ))}{\delta }y_{0}\right) v.
\end{align*}%
The eigenvalues $\lambda _{i}$ are therefore the roots of the quadratic
equation 
\begin{equation*}
\lambda _{i}^{2}+B\lambda _{i}+C=0
\end{equation*}%
where 
\begin{align*}
B& =-\frac{1}{2}(2+\delta )x_{i}-\frac{3}{2}(1+3\omega )y_{i} \\
C& =-\frac{\delta }{2}(2+3\delta )x_{i}^{2}-(2+\delta (1+3\delta
))x_{i}y_{i}+\frac{1}{2\delta }(2-11\delta -6\omega (1-\delta )+9\delta
\omega ^{2})y_{i}^{2}.
\end{align*}%
If $B>0$ and $C>0$ then both values of $\lambda _{i}$ are negative, and we
have a stable critical point. If $B<0$ and $C>0$ both values of $\lambda
_{i} $ are positive, and the critical point is unstable to perturbations. $%
C>0$ gives a saddle-point.

For points $1$ (upper branch) and $2$ (lower branch) this gives 
\begin{equation*}
B= \mp \frac{(1+\delta (2+3 \omega)-3 \omega)}{\delta \sqrt{(1+\delta) (2-3
\omega)}} \qquad \text{and} \qquad C= - \frac{(1+4 \delta-3 \omega)}{\delta
(1+\delta)}
\end{equation*}
and for points $3$ (upper branch) and $4$ (lower branch) 
\begin{equation*}
B = \pm \frac{3 (1-\omega (1+2 \delta))}{\sqrt{2 (1+\delta) (2-3 \delta
(1+\omega)-2 \delta^2 (4+3 \omega))}} \qquad \text{and} \qquad C= \frac{(1+4
\delta -3 \omega)}{\delta (1+\delta)}.
\end{equation*}

The stability of the critical points at finite distances for a universe
filled with pressureless dust are therefore, for various different values of 
$\delta $, given by

\begin{center}
\begin{tabular}{c|cc|ccc}
\label{w=0} \textbf{Critical point} & \textbf{B} & \textbf{C} & $-\frac{%
\sqrt{73}+3}{16} <\delta <-\frac{1}{4}$ & $-\frac{1}{4} <\delta<0$ & $%
0<\delta <\frac{\sqrt{73}-3}{16} $ \\ \hline
1 & $-\frac{(1+2 \delta)}{\delta \sqrt{2 (1+\delta)}}$ & $-\frac{(1+4 \delta)%
}{\delta (1+\delta)}$ & Saddle & Stable & Saddle \\ 
2 & $\frac{(1+2 \delta)}{\delta \sqrt{2 (1+\delta)}}$ & $-\frac{(1+4 \delta)%
}{\delta (1+\delta)}$ & Saddle & Unstable & Saddle \\ 
3 & $\frac{3}{\sqrt{2 (1+\delta) (2-3 \delta-8 \delta^2)}}$ & $\frac{(1+4
\delta)}{\delta (1+\delta)}$ & Stable & Saddle & Stable \\ 
4 & $-\frac{3}{\sqrt{2 (1+\delta) (2-3 \delta-8 \delta^2)}}$ & $\frac{(1+4
\delta)}{\delta (1+\delta)}$ & Unstable & Saddle & Unstable%
\end{tabular}
\end{center}

and for a universe filled with black-body radiation are given by

\begin{center}
\begin{tabular}{c|cc|c}
\label{w=1/3} \textbf{Critical point} & \textbf{B} & \textbf{C} & $-\frac{%
\sqrt{6}+1}{5} <\delta < \frac{\sqrt{6}-1}{5}$ \\ \hline
1 & $-\frac{3}{\sqrt{1+\delta}}$ & $-\frac{4}{(1+\delta)}$ & Saddle \\ 
2 & $\frac{3}{\sqrt{1+\delta}}$ & $-\frac{4}{(1+\delta)}$ & Saddle \\ 
3 & $\frac{(1-\delta)}{\sqrt{(1+\delta) (1-2\delta-5 \delta^2)}}$ & $\frac{4%
}{(1+\delta)}$ & Stable \\ 
4 & $-\frac{(1-\delta)}{\sqrt{(1+\delta) (1-2\delta-5 \delta^2)}}$ & $\frac{4%
}{(1+\delta)}$ & Unstable%
\end{tabular}
\end{center}

Values of $\delta<-\frac{\sqrt{73+66\omega+9\omega^2}+3(1+\omega)}{4 (4+3
\omega)}$ and $\delta>\frac{\sqrt{73+66\omega+9\omega^2}-3(1+\omega)}{4 (4+3
\omega)}$ have not been considered here as they lead to negative values of $%
\rho_0$ for the solution (\ref{power}).  (The reader may note the
difference here between the range of $\delta$ for which point 3 is a
stable attractor compared with the analysis of Carloni et. al.).

Point 3 lies in the $y>0$ region and so corresponds to the expanding
power--law solution (\ref{power}). It can be seen from the table above that
this solution is stable for certain ranges of $\delta $ and a saddle-point
for others. In contrast, point 4, the contracting power--law solution (\ref%
{power}), is unstable or a saddle-point. The nature of the stability of
these points and the trajectories which are attracted towards them will be
explained further in the next section.

A similar analysis can be performed for the critical points at infinity.
This time only the variable $\phi $ will be perturbed as 
\begin{equation}
\phi (t)=\phi _{i}+q(t).  \label{phipert}
\end{equation}%
The conditions for stability of the critical points are now that $r^{\prime
}>0$ and the eigenvalue $\mu $ of the linearised equation $q^{\prime }=\mu q$
satisfies $\mu <0$, in the limit $r\rightarrow 1$. If both of these
conditions are satisfied then the point is a stable attractor, if only one
is satisfied the point is a saddle-point and if neither are satisfied then
the point is repulsive.

Substituting (\ref{phipert}) into (\ref{phase4}) and linearising in $q(t)$
gives, in the limit $r\rightarrow 1$, 
\begin{align*}
q^{\prime }& =\frac{1}{4\delta ^{2}(1-r)}\left( (6(1-\delta )+\delta
^{2}(1+2\delta ))\cos \phi _{i}-3(2(1-\delta )-\delta ^{2}(1+2\delta ))\cos
3\phi _{i}-3\delta \sin \phi _{i}+9\delta \sin 3\phi \right) q \\
& \equiv \mu q.
\end{align*}%
The sign of $r^{\prime }$ in the limit $r\rightarrow 1$ can be read off from
(\ref{phase3}). The stability properties of each of the stationary points at
infinity can now be summarised in the table below

\begin{center}
\begin{tabular}{c|cccc}
\label{w=0infty} \textbf{Critical point} & $-N_1 <\delta <-\frac{1}{2}$ & $-%
\frac{1}{2}<\delta<0$ & $0<\delta<\frac{1}{4}$ & $\frac{1}{4}<\delta <N_2 $
\\ \hline
5 & Stable & Saddle & Unstable & Unstable \\ 
6 & Unstable & Saddle & Stable & Stable \\ 
7 & Unstable & Unstable & Saddle & Stable \\ 
8 & Stable & Stable & Saddle & Unstable \\ 
9 & Saddle & Stable & Stable & Saddle \\ 
10 & Saddle & Unstable & Unstable & Saddle%
\end{tabular}
\end{center}

where $N_1=\frac{\sqrt{73+66\omega+9\omega^2}+3(1+\omega)}{4 (4+3 \omega)}$
and $N_2=\frac{\sqrt{73+66\omega+9\omega^2}-3(1+\omega)}{4 (4+3 \omega)}$.

\subsubsection{Illustration of the phase plane}

\begin{figure}[tbp]
\epsfig{file=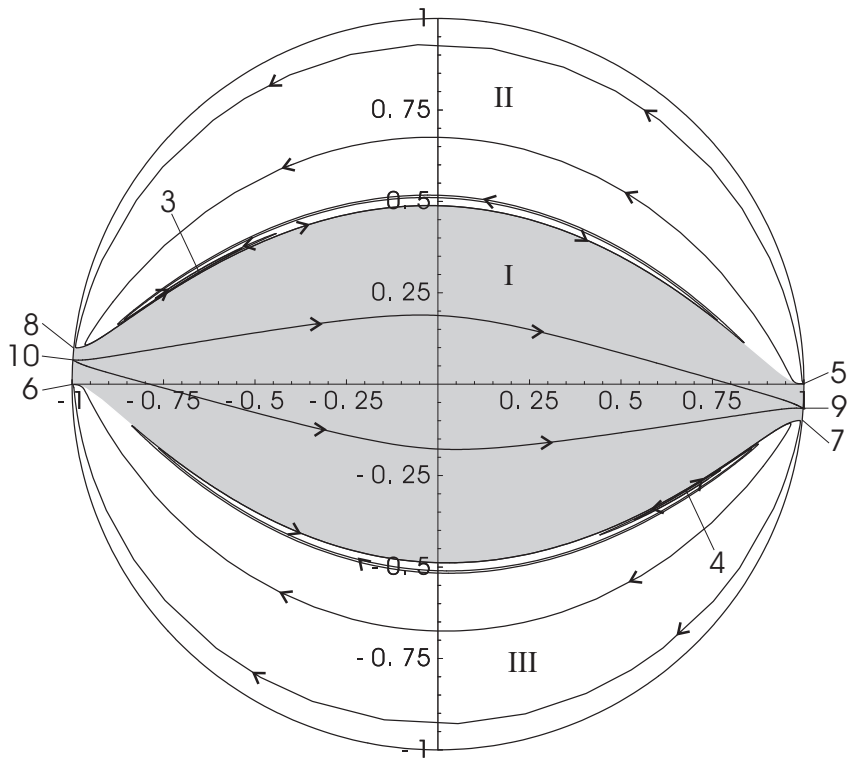,height=10cm}
\caption{\textit{Phase plane of cosmological solutions for $\protect\omega %
=0 $ and $\protect\delta =0.1$.}}
\label{d0.1}
\end{figure}

Some representative illustrations of the phase plane are now presented.
Firstly, the compactified phase plane for a universe filled with
pressureless dust, $\omega =0$, and a value of $\delta =0.1$ is shown in
figure \ref{d0.1}. 
Figure \ref{d0.1} is seen to be split into three separate regions labelled
I, II and III. The boundaries between these regions are the sub-manifolds $%
R=0$. As pointed out by Carloni et. al. the plane $R=0$ is an invariant
sub-manifold of the phase space through which trajectories cannot pass.

The equation for $R$ in a FRW universe, (\ref{R2}), can now be rewritten as 
\begin{equation}
R=\frac{16\pi \rho }{\delta R^{\delta }}((1+\delta )y^{2}+\delta (1+\delta
)xy-1).  \label{xy}
\end{equation}%
This shows that the boundary $R=0$ is given in terms of $x$ and $y$ by $%
(1+\delta )y^{2}+\delta (1+\delta )xy-1=0$ and that in region I the sign of $%
R$ must be opposite to the sign of $\delta $ in order to have a positive $%
\rho $. Similarly, in regions II and III, $R$ must have the same sign as $%
\delta $ in order to ensure a positive $\rho $.

It can be seen that regions II and III are symmetric under a rotation of $%
\pi $ and a reversal of the direction of the trajectories. As region II is
exclusively in the semi--circle $y\geqslant 0$ all trajectories confined to
this region correspond to eternally expanding (or expanding and
asymptotically static) universes. Similarly, region III is confined to the
semi--circle $y \leqslant 0$ and so all trajectories confined to this region
correspond to eternally contracting (or contracting and asymptotically
static) universes. Region I, however, spans the $y=0$ plane and so can have
trajectories which correspond to universes with both expanding and
contracting phases. In fact, it can be seen from figure \ref{d0.1} that, for 
$\delta=0.1$ all trajectories in region I are initially expanding and
eventually contracting.

It can be seen from figure \ref{d0.1} that in region I the only stable
attractors are, at early times, the expanding point 10 and at late times the
contracting point 9. (By `attractors at early times' we mean the critical
points which are approached if the trajectories are followed backwards in
time). Both of these points correspond to the solution

\begin{equation*}
a\propto t^{\delta \frac{(1+2\delta )}{(1-\delta )}}
\end{equation*}
which describes a slow evolution independent of the matter content of the
universe. Notably, region I only has stable attractor points, at both early
and late times, which have been shown to correspond to $t\rightarrow $%
constant; region I therefore does not have an asymptotic attractor when $%
t\rightarrow \infty $, for the range of $\delta $ being considered. In
region II the only stable attractors can be seen to be the static point 5 at
some early finite time, $t_{0}$, and the expanding matter-driven expansion
described by point 3 as $t\rightarrow \infty $. Conversely, in region III
the only stable attractors are the contracting point 4 for $t\rightarrow
-\infty $ and the static point 6 for $t\rightarrow t_{0}$.

\begin{figure}[tbp]
\epsfig{file=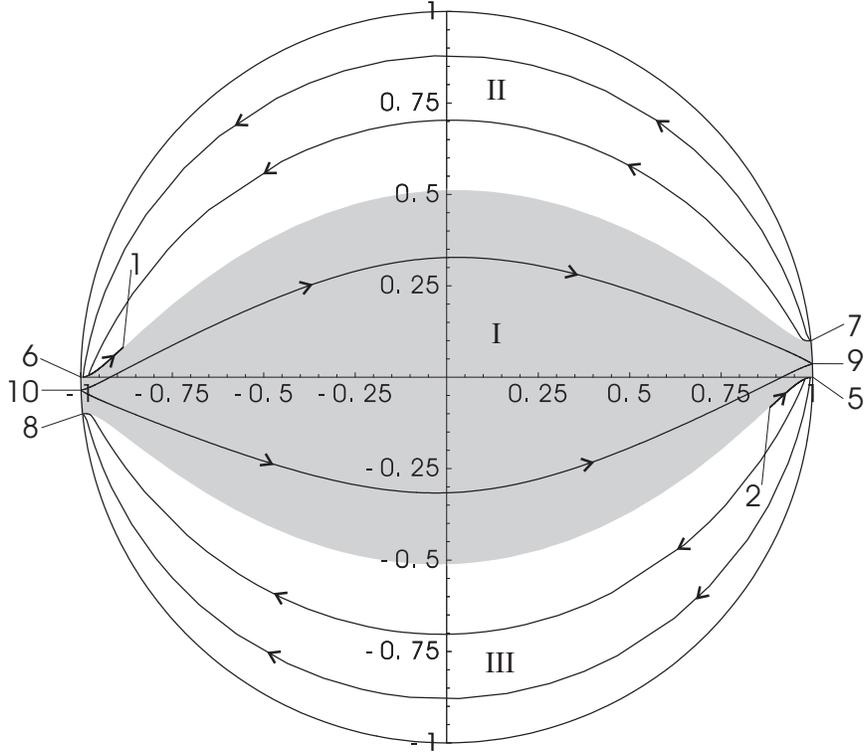,height=10cm}
\caption{\textit{Phase plane of cosmological solutions for $\protect\omega %
=0 $ and $\protect\delta =-0.1$.}}
\label{d-0.1}
\end{figure}

Figure \ref{d-0.1} shows the compactified phase plane for a universe
containing pressureless dust and having $\delta =-0.1$. 
Figure \ref{d-0.1} is split into three separate regions in a similar way to
figure \ref{d0.1}, with the boundary between the regions again corresponding
to $R=0$ and is given in terms of $x$ and $y$ by (\ref{xy}). Regions II and
III again correspond to expanding and contacting solutions, respectively.
Region I, still has point 10 as the early-time attractor and point 9 as the
late-time attractor, but now has all trajectories initially contracting and
eventually expanding. There are still no stable attractors in Region I which
correspond to regions where $t\rightarrow \infty $. Region II now has point
7 as an early-time stable attractor solution and point 1 as a late-time
stable attractor solution, corresponding to $a\rightarrow t^{\frac{1}{2}}$
as $t\rightarrow \infty $. Point 3, which was the stable attractor at late
times when $\delta =0.1$ is now no longer located in Region II and can
instead be located in region I where it is now a saddle-point in the phase
plane. Interestingly, the value of $\delta $ for which point 3 ceases to
behave as a stable attractor ($\delta =0$) is exactly the same value of $%
\delta $ at which the point moves from region II into region I; so as long
as point 3 can be located in region I, it is the late-time stable attractor
solution and as soon as it moves into region I it becomes a saddle-point. At
this same value of $\delta ,$ point 1 ceases to be a saddle-point and
becomes the late-time stable attractor for region II, so that region II
always has a stable late-time attractor where $t\rightarrow \infty $. Region
III behaves in a similar way to the description given for region II above,
under a rotation of $\pi $ and with the directions of the trajectories
reversed.

Phase planes diagrams for $\omega =0$ with values of $\delta $ other than $%
0.1$ and $-0.1$, but still within the range

\begin{equation*}
-\frac{\sqrt{73+66\omega +9\omega ^{2}}+3(1+\omega )}{4(4+3\omega )}<\delta <%
\frac{\sqrt{73+66\omega +9\omega ^{2}}-3(1+\omega )}{4(4+3\omega )},
\end{equation*}%
look qualitatively similar to those above with some of the attractor
properties of the critical points being exchanged as they pass each
other.  Notably, for $\delta <-\frac{1}{4}$ point 3 returns to region II and once
again becomes the stable late-time attractor for trajectories in that
region. The points that are the stable attractors for any particular value
of $\delta $ can be read off from the tables in the previous section.

Universes filled with perfect fluid black-body radiation also retain
qualitatively similar phase-plane diagrams to the ones above; with the
notable difference that the point 3 is always located in region II and is
always the late-time stable attractor of that region. This can be seen
directly from the Ricci scalar for the solution (\ref{power}) which is given
by 
\begin{equation*}
R=\frac{3\delta (1+\delta )}{t^{2}}
\end{equation*}%
and can be seen to have the same sign as $\delta $, for $\delta >-1$, and so
is always found in region II.

For a spatially-flat, expanding FRW universe containing black-body radiation
we therefore have that (\ref{power}) is the generic attractor as $%
t\rightarrow \infty $. Similarly, for a spatially-flat, expanding,
matter-dominated FRW universe (\ref{power}) is the attractor solution as $%
t\rightarrow \infty $; except when $-\frac{1}{4}<\delta <0$, in which case
it is point 1 ($a\propto t^{\frac{1}{2}}$).

If we require a stable period of matter domination, during which $a(t)\sim
t^{\frac{2}{3}}$, we therefore have the theoretical constraint $\delta >0$
(or $\delta <-\frac{1}{4}$). Such a period is necessary for structure to
form through gravitational collapse in the post-recombination era of the
universe's expansion.

The effect of a non--zero curvature, $\kappa \neq 0$, on the cosmological
dynamics is similar to the general relativistic case. The role of negative
curvature ($\kappa =-1$) can be deduced by noting that its effect is similar
to that of a fluid with $\omega =-1/3$. The solution (\ref{power}) is
unstable to any perturbation away from flatness and will diverge away from $%
\kappa =0$ as $t\rightarrow \infty $. This is usually referred to as the
`flatness problem' and can be seen to exist in this theory from the analysis
of Carloni et. al. \cite{Car04}.

\subsection{Physical consequences}

The modified cosmological dynamics discussed in the last section lead to
different predictions for the outcomes of physical processes, such as
primordial nucleosynthesis and CMB formation, compared to the standard
general-relativistic model. The relevant modifications to these physical
processes, and the bounds that they can impose upon the theory, will be
discussed in this section. We will use the solutions (\ref{power}) as they
have been shown to be the generic attractors as $t\rightarrow \infty $
(except for the case $-\frac{1}{4}<\delta <0$ when $\omega =0$, which has
been excluded as physically unrealistic on the grounds of structure
formation).

\subsubsection{Primordial nucleosynthesis}

We find that the temperature-time adiabat during radiation domination for
the solution (\ref{power}) is given by the exact relation 
\begin{equation}
t^{2(1+d)}=\frac{A}{T^{4}}  \label{T(t)}
\end{equation}%
where, as usual (with units $\hbar =c=1=k_{B}$), 
\begin{equation*}
\rho =\frac{g\pi ^{2}}{15}T^{4}
\end{equation*}%
where $g$ is the total number of relativistic spin states at temperature $T$%
. The constant $A$ can be determined from the generalised Friedmann equation
(\ref{Friedman2}) and is dependent on the present day value of the Ricci
scalar, through equation (\ref{chi}). (This dependence is analogous to the
dependence of scalar--tensor theories on the evolution of the non--minimally
coupled scalar, as may be expected from the relationship between these
theories \cite{Mag94}). As a first approximation we assume the universe to
have been matter dominated throughout its later history; this allows us to
write 
\begin{equation}
A=\left( \frac{45(1-2\delta )(1-2\delta -5\delta ^{2})}{32(1-\delta )g\pi
^{3}G}\right) \left( \frac{2(1+\delta )}{3H_{0}}\right) ^{2\delta }
\label{A}
\end{equation}%
where $H_{0}$ is the value of Hubble's constant today and we have used the
solution (\ref{power}) to model the evolution of $a(t)$. Adding a recent
period of accelerated expansion will refine the constant $A$, but in the
interests of brevity we exclude this from the current analysis.

As usual, the weak-interaction time is given by 
\begin{equation*}
t_{wk}\propto \frac{1}{T^{5}}.
\end{equation*}%
The freeze--out temperature, $T_{f}$, for neutron--proton kinetic
equilibrium is then defined by 
\begin{equation*}
t(T_{f})=t_{wk}(T_{f}),
\end{equation*}%
hence the freeze--out temperature in this theory, with $\delta \neq 0$, is
related to that in the general-relativistic case with $\delta =0$, $%
T_{f}^{GR}$, by 
\begin{equation}
T_{f}=C(T_{f}^{GR})^{\frac{3(1+\delta )}{(3+5\delta )}}  \label{freeze}
\end{equation}%
where 
\begin{equation}
C=\left( \frac{(1-\delta )}{(1-2\delta )(1-2\delta -5\delta ^{2})}\right) ^{%
\frac{1}{2(2+5\delta )}}\left( \frac{45}{32g\pi ^{3}G}\right) ^{\frac{\delta
(1+\delta )}{2(3+5\delta )}}\left( \frac{3H_{0}}{2(1+\delta )}\right) ^{%
\frac{\delta }{(3+5\delta }}.  \label{freeze2}
\end{equation}%
The neutron--proton ratio, $n/p$, is now determined at temperature $T$ when
the equilibrium holds by 
\begin{equation*}
\frac{n}{p}=\exp \left( -\frac{\Delta m}{T}\right) .
\end{equation*}%
where $\Delta m$ is the neutron-proton mass difference. Hence the
neutron--proton ratio at freeze-out in the $R^{1+\delta }$ early universe is
given by 
\begin{equation*}
\frac{n}{p}=\exp \left( -\frac{\Delta m}{C(T_{f}^{GR})^{1-\varepsilon }}%
\right) ,
\end{equation*}%
where 
\begin{equation*}
\varepsilon \equiv \frac{2\delta }{3+5\delta }.
\end{equation*}

The frozen--out $n/p$ ratio in the $R^{1+\delta }$ theory is given by a
power of its value in the general relativistic case, $\left( \frac{n}{p}%
\right) _{GR}\approx 1/7$, by 
\begin{equation*}
\frac{n}{p}=\left( \frac{n}{p}\right) _{GR}^{C(T_{f}^{GR})^{\varepsilon }}.
\end{equation*}%
It is seen that when $C(T_{f}^{GR})^{\varepsilon }>1$ ($\delta <0$) there is
a smaller frozen-out neutron-proton ratio that in the general-relativistic
case and consequently a lower final helium-4 abundance than in the standard
general-relativistic early universe containing the same number of
relativistic spin states. This happens because the freeze-out temperature is
lower than in general relativity. The neutrons remain in equilibrium to a
lower temperature and their slightly higher mass shifts the number balance
more towards the protons the longer they are in equilibrium. Note that a
reduction in the helium-4 abundance compared to the standard model of
general relativity is both astrophysically interesting and difficult to
achieve (all other variants like extra particle species \cite{shv, nus},
anisotropies \cite{ht, jb1, jb2, jb3}, magnetic fields \cite{mag, YM},
gravitational waves \cite{jb1, jb2, skew}, or varying $G$ \cite{bmaeda,
newt, clift}, lead to an \textit{increase} in the expansion rate and in the
final helium-4 abundance). Conversely, when $C(T_{f}^{GR})^{\varepsilon }<1$
($\delta >0$) freeze-out occurs at a higher temperature than in general
relativity and a higher final helium-4 abundance fraction results. The final
helium-4 mass fraction $Y$ is well approximated by 
\begin{equation}
Y=\frac{2n/p}{(1+n/p)}.
\end{equation}

It is now possible to constrain the value of $\delta $ using observational
abundances of the light elements. In doing this we will use the results of
Carroll and Kaplinghat \cite{Car02} who consider nucleosynthesis with a
Hubble constant parametrised by 
\begin{equation*}
H(T)=\left( \frac{T}{1MeV}\right) ^{\alpha }H_{1}.
\end{equation*}%
Our theory can be cast into this form by substituting 
\begin{equation*}
\alpha =\frac{2}{(1+\delta )}
\end{equation*}%
and 
\begin{equation*}
H_{1}=\frac{(1+\delta )}{2}A^{-\frac{1}{2(1+\delta )}}(1MeV)^{\frac{2}{%
(1+\delta )}},
\end{equation*}%
so, taking $g=43/8$, $G=6.72\times 10^{-45}MeV^{-2}$ and $H_{0}=1.51\times
10^{-39}MeV$ \cite{Ben03}, this can be rewritten as 
\begin{equation*}
H_{1}=\frac{(1+\delta )}{2}\left( \frac{7.96\times 10^{-43}(1-\delta )}{%
(1-2\delta )(1-2\delta -5\delta ^{2})}\right) ^{\frac{1}{2(1+\delta )}%
}\left( \frac{2.23\times 10^{-39}}{(1+\delta )}\right) ^{\frac{\delta }{%
(1+\delta )}}MeV.
\end{equation*}%
Carroll and Kaplinghat use the observational abundances inferred by Olive
et. al. \cite{Oli00} 
\begin{align*}
0.228\leqslant Y_{P}& \leqslant 0.248 \\
2\leqslant 10^{5}\times & \frac{D}{H}\leqslant 5 \\
1\leqslant 10^{10}\times & \frac{^{7}Li}{H}\leqslant 3
\end{align*}%
to impose the constraint 
\begin{equation*}
H_{1}=H_{c}\left( \frac{T_{c}}{MeV}\right) ^{-\alpha }
\end{equation*}%
where $H_{c}=2.6\pm 0.9\times 10^{-23}MeV$ at $T_{c}=0.2MeV$ for $%
0.5\leqslant \eta _{10}\leqslant 50$, or $H_{c}=2.0\pm 0.3\times 10^{-23}$
for $1\leqslant \eta _{10}\leqslant 10$ and $\eta _{10}$ is $10^{10}$ times
the baryon to photon ratio.

These results can now be used to impose upon $\delta$ the constraints 
\begin{equation*}
-0.017 \leqslant \delta \leqslant 0.0012,
\end{equation*}
for $0.5 \leqslant \eta_{10} \leqslant 50$, or 
\begin{equation}
-0.0064 \leqslant \delta \leqslant 0.0012,
\end{equation}
for $1 \leqslant \eta_{10} \leqslant 10$.

\subsubsection{Horizon size at matter--radiation equality}

The horizon size at the epoch of matter--radiation equality is of great
observational significance. During radiation domination cosmological
perturbations on sub--horizon scales are effectively frozen. Once matter
domination commences, however, perturbations on all scales are allowed to
grow and structure formation begins. The horizon size at matter--radiation
equality is therefore frozen into the power spectrum of perturbations and is
observable. Calculation of the horizon sizes in this theory proceeds in a
similar way to that in Brans--Dicke theory \cite{Lid98}.

In making an estimate of the horizon sizes in $R^{1+\delta }$ theory we will
use the generalised Friedmann equation, (\ref{Friedman2}), in the form 
\begin{equation}
H^{2}+\delta H\frac{\dot{R}}{R}-\frac{\delta R}{6(1+\delta )}=\frac{8\pi
G(1-\delta )}{3(1-2\delta )}\frac{R_{0}^{\delta }}{R^{\delta }}\rho .
\label{Fred}
\end{equation}
Again, we assume the form (\ref{power}) to model the evolution of the scale
factor during the epoch of matter domination. This gives the results 
\begin{align*}
a(t)& =a_{0}\left( \frac{t}{t_{0}}\right) ^{\frac{2(1+\delta )}{3}} \\
H_{0}& =\frac{2(1+\delta )}{3t_{0}} \\
\rho _{m}& =\frac{3H_{0}^{2}}{16\pi G}\frac{(1-2\delta )(2-3\delta -8\delta
^{2})}{(1-\delta )(1+\delta )^{2}}\frac{a_{0}^{3}}{a^{3}} \\
R(t)& =\frac{4(1+5\delta +4\delta ^{2})}{3t^{2}}
\end{align*}%
during the matter-dominated era. In order to simplify matters we assume the
above solutions to hold exactly from the time of matter--radiation equality
up until the present (neglecting the small residual radiation effects and
the very late time acceleration). Substituting them into (\ref{Fred}) along
with $\rho _{eq}=2\rho _{meq}$ at equality we can then solve for $H_{eq}$ to
first order in $\delta $ to find 
\begin{equation}
\frac{a_{eq}H_{eq}}{a_{0}H_{0}}\simeq \sqrt{2}\sqrt{1+z_{eq}}^{\frac{%
1-2\delta }{1+\delta }}(1-2.686\delta )+O(\delta ^{2})  \label{Heq}
\end{equation}%
where $z_{eq}$ is the redshift at matter radiation equality and $H$ has been
treated as an independent parameter. The value of $1+z_{eq}$ can now be
calculated in this theory as 
\begin{equation}
1+z_{eq}=\frac{\rho _{r0}}{\rho _{m0}}.  \label{zeq}
\end{equation}%
Taking the present day temperature of the of the microwave background as $%
T=2.728\pm 0.004K$ \cite{Fix96} gives 
\begin{equation}
\rho _{r0}=3.37\times 10^{-39}MeV^{4}  \label{rhor0}
\end{equation}%
where three families of light neutrinos have been assumed at a temperature
lower than that of the microwave background by a factor $(4/11)^{\frac{1}{3}%
} $. Using the same values for $G$ and $H_{0}$ as above we than find from
the above expression for $\rho _{m}$ that 
\begin{equation}
\rho _{m0}=2.03\times 10^{-35}\frac{(1-2\delta )(2-3\delta -8\delta ^{2})}{%
(1-\delta )(1+\delta )^{2}}MeV^{4}.  \label{rhom0}
\end{equation}%
Substituting (\ref{rhor0}), (\ref{rhom0}) and (\ref{zeq}) into (\ref{Heq})
then gives the expression for the horizon size at equality, to first order
in $\delta $, as 
\begin{equation}
\frac{a_{eq}H_{eq}}{a_{0}H_{0}}\simeq 155(1-19\delta )+O(\delta ^{2}).
\end{equation}%
This expression shows that the horizon size at matter--radiation equality
will be shifted by $\sim 1\%$ for a value of $\delta \sim 0.0005$. This
shift in horizon size should be observable in a shift of the peak of the
power spectrum of perturbations, compared to its position in the standard
general relativistic cosmology. Microwave background observations,
therefore, allow a potentially tight bound to be derived on the value of $%
\delta $. This effect is analogous to the shift of power--spectrum peaks in
Brans--Dicke theory (see e.g. \cite{Lid98}, \cite{Che99}).

A full analysis of the spectrum of perturbations in this theory requires a
knowledge of the evolution of linearised perturbations as well as a
marginalization over other parameters which can mimic this effect (e.g.
baryon density). Such a study is beyond the scope of the present work.

\section{Static and spherically--symmetric solutions}

In order to test the $R^{n}$ gravity theory in the weak-field limit by means
of the standard solar-system tests of general relativity we need to find the
analogue of the Schwarzschild metric in this generalised theory and use it
to describe the gravitational field of the Sun. In the absence of any matter
the field equations (\ref{field}) can be written as 
\begin{equation}
R_{ab}=\delta \left( \frac{R_{;}^{\ cd}}{R}-(1-\delta )\frac{R_{,}^{\
c}R_{,}^{\ d}}{R^{2}}\right) \left( g_{ac}g_{bd}+\frac{1}{2}\frac{(1+2\delta
)}{(1-\delta )}g_{ab}g_{cd}\right) .  \label{staticfield}
\end{equation}%
We find that an exact static spherically symmetric solution of these field
equations is given in Schwarzschild coordinates by the line--element 
\begin{equation}
ds^{2}=-A(r)dt^{2}+\frac{dr^{2}}{B(r)}+r^{2}(d\theta ^{2}+\sin ^{2}\theta
d\phi ^{2})  \label{Chan}
\end{equation}%
where 
\begin{align*}
A(r)& =r^{2\delta \frac{(1+2\delta )}{(1-\delta )}}+\frac{C}{r^{\frac{%
(1-4\delta )}{(1-\delta )}}} \\
B(r)& =\frac{(1-\delta )^{2}}{(1-2\delta +4\delta ^{2})(1-2\delta (1+\delta
))}\left( 1+\frac{C}{r^{\frac{(1-2\delta +4\delta ^{2})}{(1-\delta )}}}%
\right)
\end{align*}%
and $C=$ constant. This solution is conformally related to the $Q=0$ limit
of the one found by Chan, Horne and Mann for a static spherically--symmetric
space--time containing a scalar--field in a Liouville potential \cite{Cha95}%
. It reduces to Schwarzschild in the limit of general relativity: $\delta
\rightarrow 0$.

In order to evaluate whether or not this solution is physically relevant we
will proceed as follows. A dynamical systems approach will be used to
establish the asymptotic attractor solutions of the field equations (\ref%
{staticfield}). The field equations will then be perturbed around these
asymptotic attractor solutions and solved to first order in the perturbed
quantities. This linearised solution will then be treated as the physically
relevant static and spherically--symmetric weak--field limit of the field
equations (\ref{staticfield}) and compared with the exact solution (\ref%
{Chan}).

\subsection{Dynamical system}

The dynamical systems approach has already been applied to a situation of
this kind by Mignemi and Wiltshire \cite{Mig89}. We present a brief summary
of the relevant points of their work in the above notation; for a
comprehensive analysis the reader is referred to their paper.

Taking the value of sign$(R)$ from (\ref{Chan}) as sign$(-\delta (1+\delta
)/(1-2\delta (1+\delta )))$ and making a suitable choice of $\Omega _{0}$
allows the scalar-field potential (\ref{pot1}) to be written as 
\begin{equation}
V(\phi )=-\frac{3\delta ^{2}}{8\pi G(1-2\delta (1+\delta ))}\exp \left( {%
\sqrt{\frac{16\pi G}{3}}\frac{(1-\delta )}{\delta }\phi }\right) .
\label{pot}
\end{equation}

In four dimensions Mignemi and Wiltshire's choice of line--element
corresponds to 
\begin{equation}
d\bar{s}^{2}=e^{2U(\xi )}\left( -dt^{2}+\bar{r}^{4}(\xi )d\xi ^{2}\right) +%
\bar{r}^{2}(\xi )(d\theta ^{2}+\sin ^{2}\theta d\phi ^{2})  \label{metric}
\end{equation}%
which, after some manipulation, gives the field equations (\ref%
{conformalfield}) as 
\begin{align}
\zeta ^{\prime \prime }& =-\frac{2c_{1}^{2}(1-\delta )^{2}+6\delta ^{2}\eta
^{\prime }{}^{2}-24\delta ^{2}\eta ^{\prime }\zeta ^{\prime }-2(1-2\delta
-8\delta ^{2})\zeta ^{\prime }{}^{2}}{1-2\delta +4\delta ^{2}}-e^{2\zeta }
\label{wilt1} \\
\eta ^{\prime \prime }& =\frac{(1-2\delta -8\delta ^{2})(c_{1}^{2}(1-\delta
)^{2}+3\delta ^{2}\eta ^{\prime }{}^{2}-12\delta ^{2}\eta ^{\prime }\zeta
^{\prime }-(1-2\delta -8\delta ^{2})\zeta ^{\prime }{}^{2})}{3\delta
^{2}(1-2\delta +4\delta ^{2})}+\frac{(1-2\delta (1+\delta ))}{3\delta ^{2}}%
e^{2\zeta }  \label{wilt2}
\end{align}%
and 
\begin{equation}
e^{2\eta }=-\frac{(1-2\delta (1+\delta ))}{3\delta ^{2}(1-2\delta +4\delta
^{2})}\left( c_{1}^{2}(1-\delta )^{2}+3\delta ^{2}\eta ^{\prime
}{}^{2}-12\delta ^{2}\eta ^{\prime }\zeta ^{\prime }-(1-2\delta -8\delta
^{2})\zeta ^{\prime }{}^{2}+(1-2\delta +4\delta ^{2})e^{2\zeta }\right)
\label{wilt3}
\end{equation}%
where 
\begin{align*}
\zeta (\xi )& =U(\xi )+\log \bar{r}(\xi ) \\
\eta (\xi )& =-\frac{(1-2\delta (1+\delta ))}{3\delta ^{2}}U(\xi )+2\log 
\bar{r}(\xi )-\frac{(1-\delta )^{2}}{3\delta ^{2}}c_{1}\xi +\text{constant.}
\end{align*}%
Primes denote differentiation with respect to $\xi $ and $c_{1}$ is a
constant of integration.

\begin{figure}[tbp]
\epsfig{file=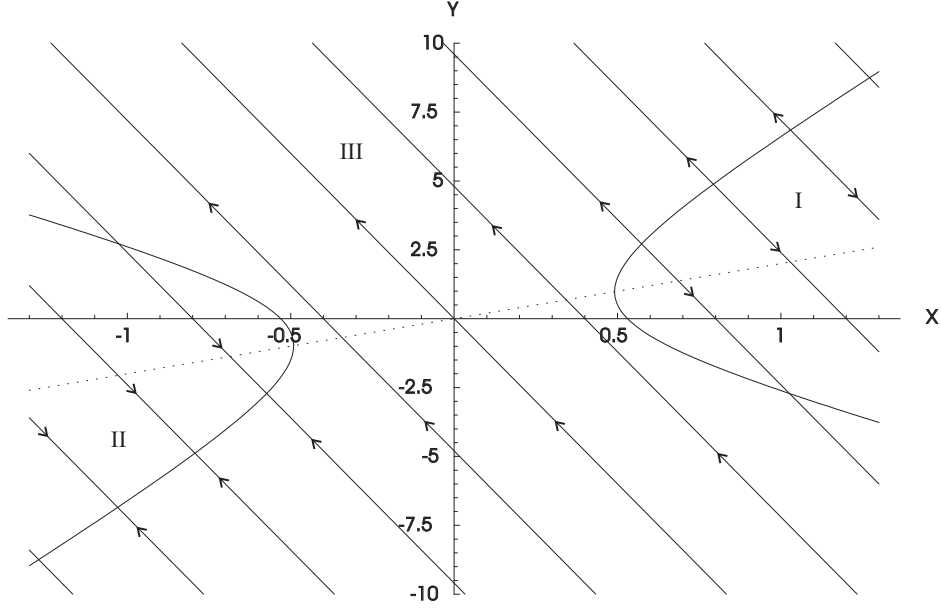,height=8cm}
\caption{\textit{The $Z=0$ plane of the phase space defined by $X$,$Y$ and $%
Z $ for $\protect\delta =0.1$ and $c_{1}=0.5$. The bold lines show the
critical points in this plane and the diagonal lines show the unphysical
trajectories confined to this plane. The dotted line is $Y=2X$ and separates
the critical points where $\protect\xi \rightarrow \infty $ from the points
where $\protect\xi \rightarrow -\infty $ }}
\label{finitephase}
\end{figure}

Defining the variables $X$, $Y$ and $Z$ by 
\begin{equation*}
X=\zeta ^{\prime }\qquad Y=\eta ^{\prime }\qquad Z=e^{\zeta }
\end{equation*}%
the field equations (\ref{wilt1}) and (\ref{wilt2}) can then be written as
the following set of first-order autonomous differential equations 
\begin{align}
X^{\prime }& =-\frac{2c_{1}^{2}(1-\delta )^{2}+6\delta ^{2}Y^{2}-24\delta
^{2}XY-2(1-2\delta -8\delta ^{2})X^{2}}{1-2\delta +4\delta ^{2}}-Z^{2}
\label{X'}
\\
\label{Y'}
Y^{\prime }& =\frac{(1-2\delta -8\delta ^{2})(c_{1}^{2}(1-\delta
)^{2}+3\delta ^{2}Y^{2}-12\delta ^{2}XY-(1-2\delta -8\delta ^{2})X^{2})}{%
3\delta ^{2}(1-2\delta +4\delta ^{2})}+\frac{(1-2\delta (1+\delta ))}{%
3\delta ^{2}}Z^{2} \\
Z^{\prime }& =XZ.  \label{Z'}
\end{align}%
(The reader should note the different definition of $Z$ here to that of
Mignemi and Wiltshire). As identified by Mignemi and Wiltshire, the only
critical points at finite values of $X$, $Y$ and $Z$ are in the plane $Z=0$
along the curve defined by 
\begin{equation*}
c_{1}^{2}(1-\delta )^{2}+3\delta ^{2}Y^{2}-12\delta ^{2}XY-(1-2\delta
-8\delta ^{2})X^{2}=0.
\end{equation*}%
These curves are shown as bold lines in figure \ref{finitephase}, together
with some sample trajectories from equations (\ref{X'})and (\ref{Y'}). From
the definition above we see that the condition $Z=0$ is equivalent to $\bar{r%
}e^{U}=0$. Whilst we do not consider trajectories confined to this plane to
be physically relevant we do consider the plot to be instructive as it gives
a picture of the behaviour of trajectories close to this surface and
displays the attractive or repulsive behaviour of the critical points, which
can be the end points for trajectories which could be considered as
physically meaningful. 
The dotted line in figure \ref{finitephase} corresponds to the line $Y=2X$
and separates two different types of critical points. The critical points
with $Y>2X$ can be seen to be repulsive to the trajectories in the $Z=0$ plane
and correspond to the limit $\xi \rightarrow -\infty $. Conversely, the
points with $Y<2X$ are attractive and correspond to the limit $\xi
\rightarrow \infty $. As $Z=\bar{r}e^{U},$ all critical points of this type
in the $Z=0$ plane correspond either to naked singularities, $\bar{r}%
\rightarrow 0$, or regular horizons, $\bar{r}\rightarrow $constant.

The two bold lines in figure \ref{finitephase} are the points at which the
surface defined by 
\begin{equation*}
c_1^2 (1-\delta)^2+3 \delta^2 Y^2 -12 \delta^2 X Y -(1-2 \delta-8\delta^2)
X^2 +(1-2 \delta+4\delta^2) Z^2 =0
\end{equation*}
crosses the $Z=0$ plane. This surface splits the phase space into three
separate regions between which trajectories cannot move. These regions are
labelled $I$, $II$ and $III$ in figure \ref{finitephase}. It can be seen
from (\ref{wilt3}) that trajectories are confined to either regions $I$ or $%
II$, for the potential defined by (\ref{pot}). If we had chosen the opposite
value of sign$(R)$ in (\ref{pot1}) then trajectories would be confined to
region $III$. We will show, however, that region $III$ does not contain
solutions with asymptotic regions in which $\bar{r} \rightarrow \infty$ and
so is of limited interest for our purposes.

\begin{figure}[tbp]
\epsfig{file=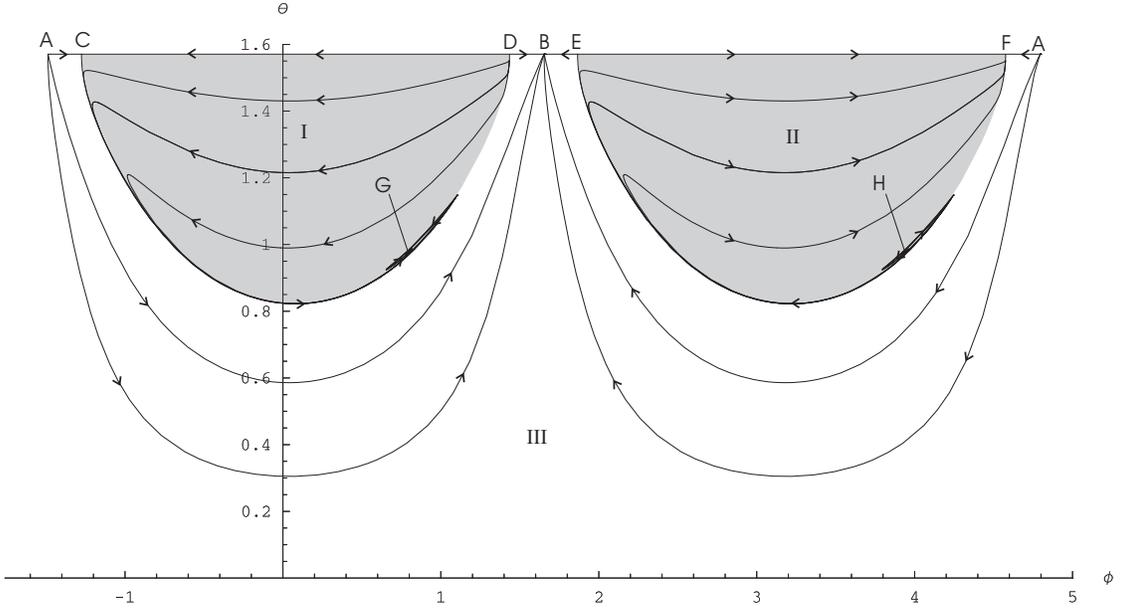,height=8cm}
\caption{\textit{The surface at infinity of the phase space defined by $X$,$%
Y $ and $Z$ for $\protect\delta =0.1$. The shaded areas show where regions $I
$ and $II$. Region $III$ is unshaded.}}
\label{infinityphase}
\end{figure}

In order to find the remaining critical points it is necessary to analyse
the sphere at infinity. This can be done by making the transformation 
\begin{equation*}
X=\rho \sin \theta \cos \phi \qquad Y=\rho \sin \theta \sin \phi \qquad
Z=\rho \cos \theta
\end{equation*}%
and taking the limit $\rho \rightarrow \infty $. The set of equations (\ref%
{X'}), (\ref{Y'}) and (\ref{Z'}) then give 
\begin{multline*}
\frac{d\theta }{d\tau }\rightarrow -\frac{\cos \theta }{24\delta
^{2}(1-2\delta +4\delta )}(6\delta ^{2}\cos \phi (3-3\delta (2-9\delta
)+(1-\delta (2+11\delta ))\cos 2\theta ) \\
-(3-3\delta (4-\delta (15-22\delta -32\delta ^{2}))+(5-\delta (20-\delta
(3+34\delta +32\delta ^{2})))\cos 2\theta )\sin \phi \\
-2(18\delta ^{2}(1-\delta (2+7\delta ))\cos 3\phi -(1-\delta (4+\delta
(9-26\delta +32\delta ^{2})))\sin 3\phi )\sin ^{2}\theta )
\end{multline*}%
and 
\begin{multline*}
\frac{d\phi }{d\tau }\rightarrow -\frac{1}{24\delta ^{2}(1-2\delta +4\delta
^{2})}(6\delta ^{2}(1-\delta (2+41\delta )-5(1-\delta (2+5\delta ))\cos
2\theta )\text{cosec}\theta \sin \phi \\
+2((1-\delta (4+\delta (9-26\delta +32\delta ^{2})))\cos 3\phi +18\delta
^{2}(1-\delta (2+7\delta ))\sin 3\phi )\sin \theta \\
-2\cos \phi (4(1-2\delta (2-\delta (3-2\delta -4\delta ^{2})))\text{cosec}%
\theta -(7-\delta (28+\delta (15-2\delta (43+128\delta ))))\sin \theta ))
\end{multline*}%
where $d\tau =\rho d\xi $. These equations can be used to plot the positions
of critical points and trajectories on the sphere at infinity. The result of
this is shown in figure \ref{infinityphase}. Once again, these trajectories
do not correspond to physical solutions in the phase space but are
illustrative of trajectories at large distances and help to show the
attractive or repulsive nature of the critical points. 
The surface at infinity has eight critical points, labelled $A$-$H$ in
figure \ref{infinityphase}. Points $A$ and $B$ are the end-points of the
trajectory that goes through the origin in figure \ref{finitephase} and are
located at 
\begin{equation*}
\theta =\frac{\pi }{2},\qquad \text{and}\qquad \phi _{1,(2)}=\cos
^{-1}\left( \frac{-6\delta ^{2}}{\sqrt{1-4\delta -12\delta ^{2}+32\delta
^{3}+100\delta ^{4}}}\right) (+\pi )
\end{equation*}%
or, in terms of the original functions in the metric (\ref{metric}), 
\begin{equation*}
\bar{r}\rightarrow (\xi -\xi _{1})^{\frac{3\delta ^{2}}{(1-2\delta -8\delta
^{2})}}\qquad \text{and}\qquad e^{U}\rightarrow (\xi -\xi _{1})^{\frac{%
3\delta ^{2}}{(1-2\delta -8\delta ^{2})}},
\end{equation*}%
where $\xi _{1}$ is a constant of integration. The points $A$ and $B$
therefore both correspond to $\xi \rightarrow \xi _{1}$ and hence to $\bar{r}%
\rightarrow 0$.

Points $C$, $D$, $E$ and $F$ are the four end points of the two curves in
figure \ref{finitephase} and therefore correspond to $\xi \rightarrow \infty$
or $-\infty$ and $\bar{r} \rightarrow 0$ or constant.

The remaining points, $G$ and $H$, are located at 
\begin{equation*}
\phi _{1,(2)}=\frac{\pi }{4}(+\pi )\qquad \text{and}\qquad \theta =\frac{1}{2%
}\cos ^{-1}\left( -\frac{1-2\delta +10\delta ^{2}}{3-6\delta +6\delta ^{2}}%
\right)
\end{equation*}%
or 
\begin{equation}
\bar{r}^{\frac{(1-2\delta +4\delta ^{2})}{(1-\delta )^{2}}}\rightarrow \pm 
\sqrt{\frac{(1-2\delta -2\delta ^{2})}{(1-2\delta +4\delta ^{2})}}\frac{1}{%
(\xi -\xi _{2})}\qquad \text{and}\qquad e^{\frac{(1-2\delta +4\delta ^{2})U}{%
3\delta ^{2}}}\rightarrow \pm \sqrt{\frac{(1-2\delta -2\delta ^{2})}{%
(1-2\delta +4\delta ^{2})}}\frac{1}{(\xi -\xi _{2})},  \label{GH}
\end{equation}%
where $\xi _{2}$ is an integration constant, the positive branch corresponds
to point $H$ and the negative branch to point $G$. These points are,
therefore, the asymptotic limit of the exact solution (\ref{Chan}) and
correspond to $\xi \rightarrow \xi _{2}$ and hence $\bar{r}\rightarrow
\infty $.

Whilst it may initially appear that trajectories are repelled from the point 
$H$, this is only the case in terms of the coordinate $\xi $. In terms of
the more physically relevant quantity $\bar{r},$ the point $H$ is an
attractor. This can be seen from the first equation in (\ref{GH}). Taking
the positive branch here it can be seen that $\bar{r}$ increases as $\xi $
decreases. So, in terms of $\bar{r}$ the points $G$ and $H$ are both
attractors, as $\bar{r}\rightarrow \infty $.

We can now see that in region $I$ all trajectories appear to start at
critical points corresponding to either $\bar{r} \rightarrow 0$ or constant
and end at point $G$ where $\bar{r} \rightarrow \infty$. Region $II$ appears
to share the same features as region $I$ with all trajectories starting
at either $\bar{r} \rightarrow 0$ or constant and ending at $H$ where $\bar{r%
} \rightarrow \infty$. Region $III$ has no critical points corresponding to $%
\bar{r} \rightarrow \infty$ and so all trajectories both begin and end on
points corresponding to $\bar{r} \rightarrow 0$ or constant.

Therefore the only solutions with an asymptotic region in which $\bar{r}%
\rightarrow \infty $ exist in regions $I$ and $II$ where the potential can
be described by equation (\ref{pot}). Furthermore, all trajectories in
regions $I$ and $II$ appear to be attracted to the solution 
\begin{equation}
ds^{2}=-\bar{r}^{\frac{6\delta ^{2}}{(1-\delta )^{2}}}dt^{2}+\frac{%
(1-2\delta +4\delta ^{2})(1-2\delta -2\delta ^{2})}{(1-\delta )^{4}}d\bar{r}%
^{2}+\bar{r}^{2}(d\theta ^{2}+\sin ^{2}\theta d\phi ^{2}),
\label{asymptotic}
\end{equation}%
which is the asymptotic behaviour of the solution found by Chan, Horne and
Mann \cite{Cha95}. We therefore conclude that all solutions with an
asymptotic region in which $\bar{r}\rightarrow \infty $ are attracted
towards the solution (\ref{asymptotic}) as $\bar{r}\rightarrow \infty $.

Rescaling the metric back to the original conformal frame we therefore
conclude that the generic asymptotic attractor solution to the field
equations, (\ref{staticfield}), is 
\begin{equation}
ds^{2}=-r^{2\delta \frac{(1+2\delta )}{(1-\delta )}}dt^{2}+\frac{(1-2\delta
+4\delta ^{2})(1-2\delta -2\delta ^{2})}{(1-\delta )^{2}}dr^{2}+r^{2}(d%
\theta ^{2}+\sin ^{2}\theta d\phi ^{2})  \label{asymptotic2}
\end{equation}%
as $r\rightarrow \infty $. It reduces to Minkowski in the $\delta
\rightarrow 0$ limit of general relativity.

\subsection{Linearised solution}

We now proceed to find the general solution, to first order in
perturbations, around the background described by (\ref{asymptotic2}).
Writing the perturbed line-element as 
\begin{equation}
ds^{2}=-r^{2\delta \frac{(1+2\delta )}{(1-\delta )}}(1+V(r))dt^{2}+\frac{%
(1-2\delta +4\delta ^{2})(1-2\delta -2\delta ^{2})}{(1-\delta )^{2}}%
(1+W(r))dr^{2}+r^{2}(d\theta ^{2}+\sin ^{2}\theta d\phi ^{2})
\label{asymptotic3}
\end{equation}%
and making no assumptions about the order of $R$ the field equations (\ref%
{staticfield}) become, up to first order in $V$ and $W$, 
\begin{multline}
\frac{\delta (1+2\delta )(1+2\delta ^{2})}{(1-\delta )^{2}r^{2}}+\frac{%
(1+2\delta ^{2})}{(1-\delta )}\frac{V^{\prime }}{r}-\frac{\delta (1-2\delta )%
}{2(1-\delta )}\frac{W^{\prime }}{r}+\frac{V^{\prime \prime }}{2}  \label{rr}
\\
=\frac{\delta (1+2\delta )}{2}\frac{R^{\prime }{}^{2}}{R^{2}}-\frac{\delta
(1+2\delta )}{2(1-\delta )}\frac{R^{\prime \prime }}{R}-\frac{3\delta }{%
4(1-\delta )}\frac{R^{\prime }}{R}V^{\prime }-\frac{\delta (1+2\delta
)(2+\delta )}{2(1-\delta )^{2}r}\frac{R^{\prime }}{R}+\frac{\delta
(1+2\delta )}{4(1-\delta )}\frac{R^{\prime }}{R}W^{\prime },
\end{multline}%
\begin{multline}
\frac{\delta (1+2\delta )(1-2\delta -2\delta ^{2})}{(1-\delta )^{2}r^{2}}-%
\frac{\delta (1+2\delta )}{(1-\delta )}\frac{V^{\prime }}{r}+\frac{(2-\delta
+2\delta ^{2})}{2(1-\delta )}\frac{W^{\prime }}{r}-\frac{V^{\prime \prime }}{%
2}  \label{tt} \\
=-\frac{3\delta }{2}\frac{R^{\prime }{}^{2}}{R^{2}}+\frac{3\delta }{%
2(1-\delta )}\frac{R^{\prime \prime }}{R}+\frac{\delta (1+2\delta )(2-\delta
+2\delta ^{2})}{2(1-\delta )^{2}r}\frac{R^{\prime }}{R}+\frac{\delta
(1+2\delta )}{4(1-\delta )}\frac{R^{\prime }}{R}V^{\prime }-\frac{3\delta }{%
4(1-\delta )}\frac{R^{\prime }}{R}W^{\prime }
\end{multline}%
and 
\begin{multline}
-\frac{2\delta (3-4\delta +2\delta ^{2}+8\delta ^{3})}{(1-\delta )^{2}r^{2}}+%
\frac{2(1-2\delta +4\delta ^{2})(1-2\delta -2\delta ^{2})}{(1-\delta
)^{2}r^{2}}W-\frac{V^{\prime }}{r}+\frac{W^{\prime }}{r}  \label{thth} \\
=-\delta (1+2\delta )\frac{R^{\prime }{}^{2}}{R^{2}}+\frac{\delta (1+2\delta
)}{(1-\delta )}\frac{R^{\prime \prime }}{R}+\frac{\delta (4-\delta +2\delta
^{2}+4\delta ^{3})}{(1-\delta )^{2}r}\frac{R^{\prime }}{R}+\frac{\delta
(1+2\delta )}{2(1-\delta )}\frac{R^{\prime }}{R}V^{\prime }-\frac{\delta
(1+2\delta )}{2(1-\delta )}\frac{R^{\prime }}{R}W^{\prime }.
\end{multline}

Expanding $R$ to first order in $V$ and $W$ gives 
\begin{equation}  \label{R}
R=-\frac{6 \delta (1+\delta)}{(1-2\delta-2 \delta^2)} \frac{1}{r^2}+R_1
\end{equation}
where 
\begin{multline}  \label{R1}
R_1= \frac{2 (1+\delta+\delta^2)}{(1-2\delta-2\delta^2)} \frac{W}{r^2} -%
\frac{2 (1-\delta) (1+2 \delta^2)}{(1- 2\delta-2\delta^2)(1-2 \delta+4
\delta^2)} \frac{V^{\prime}}{r} \\
+\frac{(1-\delta)(2-\delta+2 \delta^2)}{(1- 2\delta-2\delta^2)(1-2 \delta+4
\delta^2)} \frac{W^{\prime}}{r}- \frac{(1-\delta)^2}{(1-
2\delta-2\delta^2)(1-2 \delta+4 \delta^2)} V^{\prime\prime}.
\end{multline}

Substituting (\ref{R}) into the field equations (\ref{rr}), (\ref{tt}) and (%
\ref{thth}) and eliminating $R_1$ using (\ref{R1}) leaves 
\begin{multline*}
\frac{(1+\delta+\delta^2)(5-12\delta+12 \delta^2+4 \delta^3)}{3 (1-\delta)^2
(1+\delta)} \frac{W}{r} +\frac{(16-47\delta+76 \delta^2-34 \delta^3-16
\delta^4+32 \delta^5)}{6 (1-\delta^2) (1-2\delta+4 \delta^2)} W^{\prime} \\
-\frac{(1+\delta +7 \delta^2-19 \delta^3+44 \delta^4+20 \delta^5)}{3
(1-\delta^2) (1-2\delta+4 \delta^2)} \frac{Y}{r} -\frac{(8-15 \delta+18
\delta^2 +16 \delta^3)}{6 (1+\delta) (1-2\delta+4\delta^2)} Y^{\prime} \\
= -\frac{(1-2\delta-2\delta^2)(5-12\delta+12\delta^2+4\delta^3)}{12
(1-\delta^2) (1+\delta)} \frac{\psi}{r} -\frac{(1-2 \delta-2\delta^2)}{4
(1-\delta^2)} \psi^{\prime},
\end{multline*}
\begin{multline*}
-\frac{(1+2 \delta) (1+\delta+\delta^2) (3-4 \delta+4 \delta^2)}{3
(1-\delta)^2 (1+\delta)} \frac{W}{r} -\frac{(1+2\delta)(2-\delta+2
\delta^2)(3 -4\delta+4\delta^2)}{6 (1-\delta^2) (1-2\delta+4\delta^2)}
W^{\prime} \\
+\frac{(3-2\delta +17\delta^2 -4 \delta^3+40 \delta^2)}{3 (1-\delta^2) (1-2
\delta+4 \delta^2)} \frac{Y}{r}+ \frac{(6-\delta+2 \delta^2+20 \delta^3)}{ 6
(1+\delta) (1-2\delta+4\delta^2)} Y^{\prime} \\
= \frac{(1+2 \delta) (1-2\delta-2 \delta^2) (3-4\delta+4 \delta^2)}{12
(1-\delta)^2 (1+\delta)} \frac{\psi}{r} + \frac{(1+2 \delta)(1-2
\delta-2\delta^2)}{12 (1-\delta^2)} \psi^{\prime}.
\end{multline*}
and 
\begin{multline*}
-\frac{2 (8-8 \delta+3 \delta^2+10 \delta^3-28 \delta^4-12 \delta^5)}{3
(1-\delta^2) (1+\delta) (1-2\delta-2 \delta^2)} \frac{W}{r} - \frac{(13-22
\delta+12 \delta^2+26 \delta^3-56 \delta^4)}{3 (1-\delta^2) (1-2 \delta-2
\delta^2)(1-2 \delta +4 \delta^2)} W^{\prime} \\
+\frac{2 (4+9 \delta^2+8 \delta^3-12\delta^4)}{3 (1+\delta) (1-2
\delta-2\delta^2)(1-2\delta+4 \delta^2)} \frac{Y}{r} + \frac{(5 -4\delta-4
\delta^2+12 \delta^3)}{3 (1+\delta) (1-2 \delta-2\delta^2)(1-2\delta+4
\delta^2)} Y^{\prime} \\
= \frac{(5-4 \delta-4 \delta^2+12 \delta^3)}{6 (1-\delta)^2 (1+\delta)} 
\frac{\psi}{r} +\frac{(1+2 \delta)}{6 (1-\delta^2)} \psi^{\prime}
\end{multline*}
where $Y= r V^{\prime}$ and $\psi=r^3 R_1^{\prime}$, subject to the
constraint (\ref{R1}).

For $-\frac{(7+3 \sqrt{21})}{20}<\delta<-\frac{(7-3\sqrt{21})}{20}$ the
general solution to this first order set of coupled equations is given, in
terms of $V$ and $W$, by 
\begin{align}  \label{V}
V(r) &= c_1 V_1(r)+c_2 V_2(r)+ c_3 V_3(r) +\text{constant} \\
W(r) &= -c_1 V_1(r)+c_2 W_2(r)+c_3 W_3(r)
\end{align}
where 
\begin{align*}
V_1 &=-r^{-\frac{(1-2 \delta+4 \delta^2)}{(1-\delta)}} \\
V_2 &= \frac{ (1+2 \delta) r^{-\frac{(1-2 \delta+4 \delta^2)}{2 (1-\delta)}}%
}{2 (2-3 \delta+12 \delta^2+16 \delta^3)} \left( (1+2 \delta)^2 \sin (A \log
r) + 2 A (1-\delta) \cos(A \log r) \right) \\
W_2 &= r^{-\frac{(1-2 \delta+4 \delta^2)}{2 (1-\delta)}} \sin (A \log r) \\
V_3 &= \frac{ (1+2 \delta) r^{-\frac{(1-2 \delta+4 \delta^2)}{2 (1-\delta)}}%
}{2 (2-3 \delta+12 \delta^2+16 \delta^3)} \left( (1+2 \delta)^2 \cos (A \log
r) - 2 A (1-\delta) \sin(A \log r) \right) \\
W_3 &= r^{-\frac{(1-2 \delta+4 \delta^2)}{2 (1-\delta)}} \cos (A \log r)
\end{align*}
and 
\begin{equation*}
A= -\frac{\sqrt{7-28 \delta+36 \delta^2-16 \delta^3-80 \delta^4}}{2
(1-\delta)}.
\end{equation*}
The extra constant in (\ref{V}) is from the integration of $Y$ and can be
trivially absorbed into the definition of the time coordinate. The above
solution satisfies the constraint (\ref{R1}) without imposing any conditions
upon the arbitrary constants $c_1$, $c_2$ and $c_3$.

It can be seen by direct comparison that the constant $c_1$ is linearly
related to the constant $C$ in (\ref{Chan}) by a factor that is a function
of $\delta$ only. The constants $c_2$ and $c_3$ correspond to two new
oscillating modes.

\subsection{Physical consequences}

In order to calculate the classical tests of metric theories of gravity
(i.e. bending and time-delay of light rays and the perihelion precession of
Mercury) we require the static and spherically symmetric solution to the
field equations (\ref{field}). Due to the complicated form of these
equations we are unable to find the general solution; instead we propose to
use the first--order solution around the generic attractor as $r\rightarrow
\infty $. This method should be applicable to gravitational experiments
performed in the solar system as the gravitational field in this region can
be considered weak and we will be considering experiments performed at large 
$r$ (in terms of the Schwarzschild radius of the massive objects in the
system). To this end we will use the solution found at the end of the
previous subsection. We choose to arbitrarily set the constants $c_{2}$ and $%
c_{3}$ to zero - this removes the oscillatory parts of the solution, and
hence ensures that the gravitational force is always attractive. This
considerable simplification of the solution also allows a straightforward
calculation of both null and timelike geodesics which can be used to compute
the outcomes of the classical tests in this theory.

\subsubsection{Solution in isotropic coordinates}

Having removed the oscillatory parts of the solution we are left with the
part corresponding to the exact solution (\ref{Chan}). Making the coordinate
transformation 
\begin{equation*}
r^{(1-2 \delta+4 \delta^2)/(1-d)}= \left(1-\frac{C}{4 \hat{r}^{\sqrt{\frac{%
(1-2 \delta+4 \delta^2)}{(1-2 \delta-2 \delta^2)}}}} \right)^2 \hat{r}^{%
\sqrt{\frac{(1-2 \delta+4 \delta^2)}{(1-2 \delta-2 \delta^2)}}}
\end{equation*}
the solution (\ref{Chan}) can be transformed into the isotropic coordinate
system 
\begin{equation}  \label{iso}
ds^2=-A(\hat{r}) d t^2+B(\hat{r}) ( d\hat{r}^2+ \hat{r}^2 (d
\theta^2+\sin^2\theta d \phi^2))
\end{equation}
where 
\begin{equation*}
A(\hat{r})= \hat{r}^{\frac{2 \delta (1+2 \delta)}{\sqrt{(1-2
\delta-2\delta^2)(1-2\delta+4 \delta^2)}}} \left(1+\frac{C}{4 \hat{r}^{\sqrt{%
\frac{(1-2 \delta+4 \delta^2)}{(1-2 \delta-2 \delta^2)}}}} \right)^2 \left(1-%
\frac{C}{4 \hat{r}^{\sqrt{\frac{(1-2 \delta+4 \delta^2)}{(1-2 \delta-2
\delta^2)}}}} \right)^{-\frac{2 (1+4 \delta)}{(1-2 \delta+4 \delta^2)}}
\end{equation*}
and 
\begin{equation*}
B(\hat{r})= \hat{r}^{-2+2\frac{(1-\delta)}{\sqrt{(1-2
\delta-2\delta^2)(1-2\delta+4 \delta^2)}}} \left(1-\frac{C}{4 \hat{r}^{\sqrt{%
\frac{(1-2 \delta+4 \delta^2)}{(1-2 \delta-2 \delta^2)}}}} \right)^{\frac{4
(1-\delta)}{(1-2 \delta+4 \delta^2)}},
\end{equation*}
which is, to linear order in $C$, 
\begin{equation*}
A(\hat{r})= \hat{r}^{\frac{2 \delta (1+2 \delta)}{\sqrt{(1-2
\delta-2\delta^2)(1-2\delta+4 \delta^2)}}} \left(1+ \frac{(1-\delta) (1-2
\delta)}{(1-2 \delta+4 \delta^2)}\frac{C}{\hat{r}^{\sqrt{\frac{(1-2 \delta+4
\delta^2)}{(1-2 \delta-2 \delta^2)}}}} \right)
\end{equation*}
and 
\begin{equation*}
B(\hat{r})= \hat{r}^{-2+2\frac{(1-\delta)}{\sqrt{(1-2
\delta-2\delta^2)(1-2\delta+4 \delta^2)}}} \left(1- \frac{(1-\delta)}{(1-2
\delta+4 \delta^2)}\frac{C}{\hat{r}^{\sqrt{\frac{(1-2 \delta+4 \delta^2)}{%
(1-2 \delta-2 \delta^2)}}}} \right).
\end{equation*}

\subsubsection{Newtonian limit}

We first investigate the Newtonian limit of the geodesic equation in order
to set the constant $C$ in the solution (\ref{iso}) above. As usual, we have 
\begin{equation*}
\Phi _{,\mu }=\Gamma _{\;00}^{\mu }
\end{equation*}%
where $\Phi $ is the Newtonian gravitational potential. Substituting in the
isotropic metric (\ref{iso}) this gives 
\begin{align}
\nabla \Phi & =\frac{\nabla A(\hat{r})}{2B(\hat{r})}  \label{motion} \\
& =\frac{\delta (1+2\delta )\hat{r}^{1-2\sqrt{\frac{1-2\delta -2\delta ^{2}}{%
1-2\delta +4\delta ^{2}}}}}{\sqrt{(1-2\delta -2\delta ^{2})(1-2\delta
+4\delta ^{2})}}-\frac{(1-\delta )(1-8\delta +4\delta ^{2})C\hat{r}^{1-\frac{%
3(1-2\delta )}{\sqrt{(1-2\delta -2\delta ^{2})(1-2\delta +4\delta ^{2})}}}}{2%
\sqrt{(1-2\delta -2\delta ^{2})(1-2\delta +4\delta ^{2})^{3}}}+O(C^{2}).
\end{align}

The second term in the expression goes as $\sim \hat{r}^{-2+O(\delta ^{2})}$
and so corresponds to the Newtonian part of the gravitational force. The
first term, however, goes as $\sim \hat{r}^{-1+O(\delta ^{2})}$ and has no
Newtonian counterpart. In order for the Newtonian part to dominate over the
non-Newtonian part we must impose upon $\delta $ the requirement that it is
at most 
\begin{equation*}
\delta \sim O\left( \frac{C}{r}\right) .
\end{equation*}%
If $\delta $ were larger than this then the non-Newtonian part of the
potential would dominate over the Newtonian part, which is clearly
unacceptable at scales over which the Newtonian potential has been measured
and shown to be accurate.

This requirement upon the order of magnitude of $\delta$ allows (\ref{motion}%
) to be written 
\begin{equation}  \label{force}
\nabla \Phi = \frac{\delta}{\hat{r}^{1+O(C^2)}}-\frac{C}{2 \hat{r}^{2+O(C^2)}%
}+O(C^2)
\end{equation}
where expansions in $C$ have been carried out separately in the coefficients
and the powers of $\hat{r}$ of the two terms.

Comparison of (\ref{force}) with the Newtonian force law 
\begin{equation*}
\nabla \Phi_N=\frac{G m}{r^2}
\end{equation*}
allows the value of $C$ to be read off as 
\begin{equation*}
C=-2 G m +O(\delta).
\end{equation*}

\subsubsection{Post--Newtonian limit}

We now wish to calculate, to post--Newtonian order, the equations of motion
for test particles in the metric (\ref{iso}). The geodesic equation can be
written in its usual form 
\begin{equation*}
\frac{d^{2}x^{\mu }}{d\lambda ^{2}}+\Gamma _{\;ij}^{\mu }\frac{dx^{i}}{%
d\lambda }\frac{dx^{j}}{d\lambda }=0,
\end{equation*}%
where $\lambda $ can be taken as proper time for a timelike geodesic or as
an affine parameter for a null geodesic. In terms of coordinate time this
can be written 
\begin{equation}
\frac{d^{2}x^{\mu }}{dt^{2}}+\left( \Gamma _{\;ij}^{\mu }-\Gamma _{\;ij}^{0}%
\frac{dx^{\mu }}{dt}\right) \frac{dx^{i}}{dt}\frac{dx^{j}}{dt}=0.
\label{motion2}
\end{equation}%
We also have the integral 
\begin{equation}
g_{ij}\frac{dx^{i}}{dt}\frac{dx^{j}}{dt}=S  \label{motion3}
\end{equation}%
where $S=-1$ for particles and $0$ for photons.

Substituting (\ref{iso}) into (\ref{motion2}) and (\ref{motion3}) gives, to
the relevant order, the equations of motion 
\begin{multline}
\frac{d^{2}\mathbf{x}}{dt^{2}}=-\frac{Gm}{r^{2}}\left( 1+\left\vert \frac{d%
\mathbf{x}}{dt}\right\vert ^{2}\right) \mathbf{e_{r}}+4\frac{G^{2}m^{2}}{%
r^{3}}\mathbf{e_{r}}+4\frac{Gm}{r^{2}}\mathbf{e_{r}}\cdot \frac{d\mathbf{x}}{%
dt}\frac{d\mathbf{x}}{dt}  \label{geo1} \\
-\frac{\delta }{r}\left( 1-\left\vert \frac{d\mathbf{x}}{dt}\right\vert
^{2}\right) \mathbf{e_{r}}-4\frac{\delta ^{2}}{r}\mathbf{e_{r}}+\delta \frac{%
Gm}{r^{2}}\mathbf{e_{r}}+O(G^{3}m^{3})
\end{multline}%
and 
\begin{equation}
\left\vert \frac{d\mathbf{x}}{dt}\right\vert ^{2}=1-4\frac{Gm}{r}+\frac{S}{%
r^{2\delta }}-2S\frac{Gm}{r^{1+2\delta }}+O(G^{2}m^{2}).  \label{geo2}
\end{equation}%
(In the interests of concision we have excluded the $O(\delta ^{2})$ terms
from the powers of $r$, the reader should regard them as being there
implicitly). The first three terms in equation (\ref{geo1}) are identical to
their general-relativistic counterparts. The next two terms are completely
new and have no counterparts in general relativity. The last term in
equation (\ref{geo1}) can be removed by rescaling the mass term by $%
m\rightarrow m(1+\delta )$; this has no effect on the Newtonian limit of the
geodesic equation as any term $Gm\delta $ is of post--Newtonian order.

\subsubsection{The bending of light and time delay of radio signals}

From equation (\ref{geo2}) it can be seen that the solution for null
geodesics, to zeroth order, is a straight line that can be parametrised by 
\begin{equation*}
\mathbf{x}=\mathbf{n}(t-t_{0})
\end{equation*}%
where $\mathbf{n}\cdot \mathbf{n}=1$. Considering a small departure from the
zeroth order solution we can write 
\begin{equation*}
\mathbf{x}=\mathbf{n}(t-t_{0})+\mathbf{x}_{1}
\end{equation*}%
where $\mathbf{x}_{1}$ is small. To first order, the equations of motion (%
\ref{geo1}) and (\ref{geo2}) then become 
\begin{equation}
\frac{d^{2}\mathbf{x}}{dt^{2}}=-2\frac{Gm}{r^{2}}\mathbf{e_{r}}+4\frac{Gm}{%
r^{2}}(\mathbf{n}\cdot \mathbf{e_{r}})\mathbf{n}  \label{geo3}
\end{equation}%
and 
\begin{equation}
\mathbf{n}\cdot \frac{d\mathbf{x}}{dt}=-2\frac{Gm}{r}.  \label{geo4}
\end{equation}

Equations (\ref{geo3}) and (\ref{geo4}) can be seen to be identical to the
first-order equations of motions for photons in general relativity. We
therefore conclude that any observations involving the motion of photons in
a stationary and spherically symmetric weak field situation cannot tell any
difference between general relativity and this $R^{1+\delta }$ theory, to
first post--Newtonian order. This includes the classical light bending and
time delay tests which should measure the post-Newtonian parameter $\gamma $ to be one
in this theory, as in general relativity.

\subsubsection{Perihelion precession}

In calculating the perihelion precession of a test particle in the geometry (%
\ref{iso}) it is convenient to use the standard procedures for computing the
perturbations of orbital elements (see \cite{Sma53} and \cite{Rob68}). In
the notation of Robertson and Noonan \cite{Rob68} the measured rate of
change of the perihelion in geocentric coordinates is given by 
\begin{equation}
\frac{d\tilde{\omega}}{dt}=-\frac{p\mathcal{R}}{he}\cos \phi +\frac{\mathcal{%
J}(p+r)}{he}\sin \phi  \label{per}
\end{equation}%
where $p$ is the semi--latus rectum of the orbit, $h$ is the
angular--momentum per unit mass, $e$ is the eccentricity and $\mathcal{R}$
and $\mathcal{J}$ are the components of the acceleration in radial and
normal to radial directions in the orbital plane, respectively. The radial
coordinate, $r$, is defined by 
\begin{equation}
r\equiv \frac{p}{(1+e\cos \phi )}  \label{r}
\end{equation}%
and $\phi $ is the angle measured from the perihelion. We have, as usual,
the additional relations 
\begin{equation*}
p=a(1-e^{2})
\end{equation*}%
and 
\begin{equation}
h\equiv \sqrt{Gmp}\equiv r^{2}\frac{d\phi }{dt}.  \label{dr}
\end{equation}

From (\ref{geo1}), the components of the acceleration can be read off as 
\begin{equation}
\mathcal{R}=-\frac{Gm}{r^{2}} -\frac{Gm}{r^2} v^2 +4\frac{Gm}{r^{2}}v_{\mathcal{R}}^{2}+4\frac{%
G^{2}m^{2}}{r^{3}}-\frac{\delta }{r}+\frac{\delta }{r}v^{2}-4\frac{\delta
^{2}}{r}  \label{Raccel}
\end{equation}%
and 
\begin{equation}
\mathcal{J}=4\frac{Gm}{r^{2}}v_{\mathcal{R}}v_{\mathcal{J}}  \label{Jaccel}
\end{equation}%
where we now have the radial and normal-to-radial components of the velocity
as 
\begin{align*}
v_{\mathcal{R}}& =\frac{eh}{p}\sin \phi \\
v_{\mathcal{J}}& =\frac{h}{p}(1+e\cos \phi )
\end{align*}%
and $v^{2}=v_{\mathcal{R}}^{2}+v_{\mathcal{J}}^{2}$. In writing (\ref{Raccel}%
), the last term of (\ref{geo1}) has been absorbed by a rescaling of $m$, as
mentioned above.

The expressions (\ref{Raccel}) and (\ref{Jaccel}) can now be substituted
into (\ref{per}) and integrated from $\phi=0$ to $2 \pi$, using (\ref{r})
and (\ref{dr}) to write $r$ and $dr$ in terms of $\phi$ and $d\phi$. The
perihelion precession per orbit is then given, to post--Newtonian accuracy,
by the expression 
\begin{equation}  \label{precession}
\Delta \tilde{\omega} = \frac{6 \pi G m}{a(1-e^2)}-\frac{2\pi\delta}{e^2}
\left( e^2-1-\frac{(1+4 \delta) a (1-e^2)}{Gm} \right).
\end{equation}
The first term in (\ref{precession}) is clearly the standard general
relativistic expression. The second term is new and contributes to leading
order the term 
\begin{equation*}
\frac{2 \pi a}{Gm} \left( \frac{1-e^2}{e^2} \right) \delta.
\end{equation*}

Comparing the prediction (\ref{precession}) with observation is a
non-trivial matter. The above prediction is the highly idealised precession
expected for a timelike geodesic in the geometry described by (\ref{iso}).
If we assume that the geometry (\ref{iso}) is a good approximation to the
weak field for a static Schwarzschild--like mass then it is not trivial to
assume that the timelike geodesics used to calculate the rate of perihelion
precession (\ref{precession}) are the paths that material objects will
follow. Whilst we are assured from the generalised Bianchi identities \cite%
{Mag94} of the covariant conservation of energy--momentum, $%
T_{\;\;;b}^{ab}=0 $, and hence the geodesic motion of an ideal fluid of
pressureless dust, $u^{i}u_{\;;i}^{j}=0$, this does not ensure the geodesic
motion of extended bodies. This deviation from geodesic motion is known as
the Nordvedt effect \cite{Nor68} and, whilst being zero for general
relativity, is generally non--zero for extended theories of gravity. From
the analysis so far it is also not clear how orbiting matter and other
nearby sources (other than the central mass) will contribute to the geometry
(\ref{iso}).

In order to make a prediction for a physical system such as the solar
system, and in the interests of brevity, some assumptions must be made. It
is firstly assumed that the geometry of space--time in the solar system can
be considered, to first approximation, as static and spherically symmetric.
It is then assumed that this geometry is determined by the Sun, which can be
treated as a point-like Schwarzschild mass at the origin, and is isolated
from the effects of matter outside the solar system and from the background
cosmology. It is also assumed that the Nordvedt effect is negligible and
that extended massive bodies, such as planets, follow the same timelike
geodesics of the background geometry as neutral test particles.

In comparing with observation it is useful to recast (\ref{precession}) in
the form 
\begin{equation*}
\Delta \tilde{\omega}=\frac{6\pi Gm}{a(1-e^{2})}\lambda
\end{equation*}%
where 
\begin{equation*}
\lambda =1+\frac{a^{2}(1-e^{2})^{2}}{3G^{2}m^{2}e^{2}}\delta .
\end{equation*}%
This allows for easy comparison with results which have been used to
constrain the standard post-Newtonian parameters, for which 
\begin{equation*}
\lambda =\frac{1}{3}(2+2\gamma -\beta ).
\end{equation*}%
The observational determination of the perihelion precession of Mercury is
not clear cut and is subject to a number of uncertainties; most notably the
quadrupole moment of the Sun (see e.g. \cite{Pir03}). We choose to use the
result of Shapiro et. al. \cite{Sha76} 
\begin{equation}
\lambda =1.003\pm 0.005  \label{Shapiro}
\end{equation}%
which for standard values of $a$, $e$ and $m$ \cite{All63} gives us the
constraint 
\begin{equation}
\delta =2.7\pm 4.5\times 10^{-19}.
\end{equation}%
In deriving (\ref{Shapiro}) the quadrupole moment of the Sun was assumed to
correspond to uniform rotation.  For more modern estimates of the
anomalous perihelion advance of Mercury see \cite{Pir03}. 

\section{Conclusions}

We have considered here the modification to the gravitational Lagrangian $%
R\rightarrow R^{1+\delta }$, where $\delta $ is a small rational number. By
considering the idealised Friedmann--Robertson--Walker cosmology and the
static and spherically symmetric weak field situations we have been able to
determine suitable solutions to the field equations which we have used to
make predictions of the consequences of this gravity theory for
astrophysical processes. These predictions have been compared to
observations to derive a number of bounds on the value of $\delta $.

Firstly, we showed that for a spatially-flat, matter-dominated universe the
attractor solution for the scale factor as $t\rightarrow \infty $ is of the
form $a(t)\propto t^{\frac{1}{2}}$ if\ $-\frac{1}{4}<\delta <0$. This is
unacceptable as sub--horizon scale density perturbations do not grow in a
universe described by a scale factor of this form. We therefore have the
constraints 
\begin{equation}
\delta \geqslant 0\qquad (\text{or}\quad \delta <-1/4),
\end{equation}%
in which case the attractor solution for the scale factor as $t\rightarrow
\infty $ changes to that of the exact solution $a(t)\propto t^{\frac{%
2(1+\delta )}{3}}$.

Secondly, we showed that the modified expansion rate during primordial
nucleosynthesis alters the predicted abundances of light elements in the
universe. Using the inferred observational abundances of Olive et. al. \cite%
{Oli00} we were able to impose upon $\delta $ the constraints 
\begin{equation}
-0.017\leqslant \delta \leqslant 0.0012,
\end{equation}%
for $0.5\leqslant \eta _{10}\leqslant 50$, or 
\begin{equation}
-0.0064\leqslant \delta \leqslant 0.0012,
\end{equation}%
for $1\leqslant \eta _{10}\leqslant 10$.

Next, we considered the horizon size at the time of matter--radiation
equality. After showing that the horizon size is different in this theory to
its counterpart in general-relativistic cosmology we discussed the
implications for microwave background observations. This argument runs in
parallel to that of Liddle et. al. for the Brans--Dicke cosmology \cite%
{Lid98}. The horizon size at matter--radiation equality will be shifted by $%
\sim 1\%$ for a value of $\delta \sim 0.0005.$

Finally, we investigated the static and spherically symmetric weak--field
situation. We calculated the null and timelike geodesics of the space--time
to post--Newtonian accuracy. We then showed that null geodesics are, to the
required accuracy, identical in this theory to those in the Schwarzschild
solution of general relativity. The light bending and radio time--delay
tests should, therefore, yield the same results as in general relativity, to
the required order.

Our prediction for the perihelion precession of Mercury gave us our tightest
bounds on $\delta $. Assuming that Mercury follows timelike geodesics of the
space--time we used the results of Shapiro et. al. \cite{Sha76} to impose
upon $\delta $ the constraint 
\begin{equation}
\delta =2.7\pm 4.5\times 10^{-19}.
\end{equation}%
This constraint is due to the unusual feature of the static and
spherically--symmetric space--time that as $r\rightarrow \infty $ it is
asymptotically attracted to a form that is not Minkowski spacetime, but
reduces to Minkowski spacetime as $\delta \rightarrow 0$.

Combining the above results we therefore have that $\delta $ should be
constrained to lie within the range 
\begin{equation}
0\leq \delta <7.2\times 10^{-19}.
\end{equation}
This is a remarkably strong observational constraint upon deviations of this kind from
general relativity.

\section{Acknowledgements}

We would like to thank Robert Scherrer, David Wiltshire and Spiros
Cotsakis for helpful comments and suggestions.  TC is supported by the PPARC.

\end{document}